\documentclass[12pt,preprint]{aastex}
\usepackage{lscape} 
\newfont{\rten}{cmr10} 
\def \arcdeg{\hbox{$^\circ$}}

\begin{document}

\title{Chaotic exchange of solid material between planetary systems: implications for lithopanspermia}

\author{Edward Belbruno\altaffilmark{a,b}, Amaya Moro-Mart\'{\i}n\altaffilmark{c,b}, Renu Malhotra\altaffilmark{d}, Dmitry Savransky\altaffilmark{e,f}}
\altaffiltext{1}{Courant Institute of Mathematical Sciences, New York University.}
\altaffiltext{2}{Department of Astrophysical Sciences, Princeton University.} 
\altaffiltext{3}{Centro de Astrobiolog\'{\i}a, INTA-CSIC, Spain.}
\altaffiltext{4}{Department of Planetary Sciences, University of Arizona.}
\altaffiltext{5}{Mechanical and Aerospace Engineering, Princeton University.}
\altaffiltext{6}{Lawrence Livermore National Laboratory.}

\begin{abstract}

We examine a low energy mechanism for the transfer of meteoroids between two planetary systems embedded in a star cluster using quasi-parabolic orbits of minimal energy. Using Monte Carlo simulations, we find that the exchange of meteoroids could have been significantly more efficient than previously estimated. Our study is relevant to astrobiology as it addresses whether life on Earth could have been transferred to other planetary systems in the solar system's birth cluster and whether life on Earth could have been transferred here from beyond the solar system. In the solar system, the timescale over which solid material was delivered to the region from where it could be transferred via this mechanism likely extended to several hundred million years (as indicated by the 3.8-4.0 Ga epoch of the Late Heavy Bombardment). This timescale could have overlapped with the lifetime of the Solar birth cluster ($\sim$ 100--500 Myr). Therefore, we conclude that lithopanspermia is an open possibility if life had an early start. Adopting parameters from the minimum mass solar nebula, considering a range of planetesimal size distributions derived from observations of asteroids and Kuiper Belt Objects and theoretical coagulation models, and taking into account Oort Cloud formation models, the expected number of bodies with mass $>$ 10 kg that could have been transferred between the Sun and its nearest cluster neighbor could be of the order of $10^{14}$--3$\cdot$$10^{16}$, with transfer timescales of 10s Myr. We estimate that of the order of $3 \cdot 10^{8} \cdot l{\rm (km)}$ could potentially be life-bearing, where $l$ is the depth of the Earth crust in km that was ejected as the result of the early bombardment. 

\end{abstract}

\keywords{Extrasolar Planets - Interplanetary Dust - Interstellar Meteorites - Lithopanspermia.}

\noindent {\footnotesize }

\noindent {\footnotesize Accepted by Astrobiology. Submitted: Sep. 21, 2011. Accepted: May 2, 2012.}

\noindent {\footnotesize Corresponding author: Amaya Moro-Mart\'{\i}n, Centro de Astrobiolog\'{\i}a, INTA, Carretera a Ajalvir km. 4, 
28850 Torrej\'on de Ardoz, Madrid, Spain. Tel. +34 91 520 6420; fax: +34 91 520 1621; email: amaya@cab.inta-csic.es.

\section{Introduction}
\label{sec:intro}

From the collection of thousands of meteorites found on Earth, approximately one hundred have been identified as having a Martian origin, and more than 46 kg of rocks have a lunar origin.  The study of the dynamical evolution of these meteorites agrees well with the cosmic ray exposure time and with their frequency of landing on Earth. A handful of meteorites has also been identified on the Moon and Mars (e.g. McSween 1976; Schr{\"o}der et al. 2008).  These findings, together with dynamical simulations (Gladman, 1997; Dones et al. 1999), indicate that meteorites are exchanged among the terrestrial planets of our solar system at a measurable level. Sufficiently large rocks may protect dormant microorganisms from ionizing radiation and from the hazards of the impact at landing. Laboratory experiments have confirmed that several microorganisms embedded in martian-like rocks could survive under shock pressures similar to those suffered by martian meteorites upon impact ejection (St\"offler et al. 2007; Horneck et al. 2008). Other studies have also shown that microorganisms in a liquid, bacteria and yeast spores can survive impacts with shock pressures of the order of GPas (Burchell et al. 2004; Willis et al. 2006; Hazell et al., 2010; Meyer et al. 2011). Therefore, it is of interest to consider that the exchange of microorganisms living inside rocks could take place among the solar system planets, a phenomenon known as {\it lithopanspermia}. Under this scenario, life on Earth could potentially spread to other moons and planets within our solar system; conversely life on Earth could have an origin elsewhere in our solar system. 

Melosh (2003) investigated quantitatively the probability of transfer taking place between the stars in the solar local neighborhood. He found that even though numerical simulations show that up to one-third of all the meteorites originating from the terrestrial planets are ejected out of the solar system  by gravitational encounters with Jupiter and Saturn, the probability of landing on a terrestrial planet of a neighboring planetary system is extremely low because of the high relative velocities of the stars and the low stellar densities: after the heavy bombardment, he estimates that only one or two rocks originating from the surface of one of the terrestrial planets may have been temporarily captured into another planetary system, with a 10$^{-4}$ probability of landing in a terrestrial planet. Therefore, he concluded that lithopanspermia among the current solar neighbors is ``overwhelmingly unlikely''.  

In a subsequent paper, Adams \& Spergel (2005) pointed out that the majority of stars, including the Sun, are born in stellar clusters where the probability of transfer would be higher due the larger stellar densities and smaller relative velocities compared to those for field stars (including the current solar neighborhood). The dispersal time of the clusters and timescale for planet formation are comparable:  the former is approximately $T = 2.3 (M_{\rm cluster}/M_\odot)^{0.6}$ Myr = (135--$535)$ Myr (for a cluster of 1000--10000 members - Adams 2010), while the latter is of order 100 Myr for terrestrial planets. Therefore, it could be possible that rocky material be transferred before the cluster disperses. Adam $\&$ Spergel (2005) estimated the probability of transfer of meteoroids between planetary systems within a cluster by using Monte Carlo simulations. To increase the capture cross-section they assumed that the stars are in binary systems. They found that clusters of  30--1000 members could experience O(10$^9$)--O(10$^{12}$) capture events among their binary members. Adopting typical ejection speeds of $\sim$5 km/s, and the number of rocky ejecta (of mass $> 10$ kg) per system of $N_R\sim10^{16}$, they found that the expected number of successful lithopanspermia events per cluster is $\sim$ 10$^{-3}$; for lower ejection speeds, $\sim$ 2 km/s, this number is 1--2. These latter estimates are relevant to the exchange of biologically active material. Valtonen et al. (2009) have also studied the exchange of solids between stars in the solar birth cluster and its enhanced capture probability compared to the exchange of solids between field stars. They concluded that approximately 10$^{2\pm2}$ bodies with sizes larger than $\sim$ 20 cm may have been exchanged between the cluster stars (compared to 10$^{-8}$ between field stars). 

Because there is a significant increase in the number of possible transfer events with decreased ejection velocity, it is of interest to study a very low energy mechanism with velocities significantly smaller than those considered in Adams \& Spergel (2005). This mechanism was described by Belbruno (2004) in the mathematical context of a class of nearly parabolic trajectories in the restricted three-body problem. The escape velocities of these parabolic-type trajectories are very low ($\sim$ 0.1 km/s), substantially smaller than the mean relative velocity of stars in the cluster, and the meteoroid escapes the planetary system by slowly meandering away. This process of ``weak escape'' is chaotic in nature.  Weak escape is a transitional motion between capture and escape. For it to occur, the trajectory of the meteoroid must pass near the largest planet in the system.  ``Weak capture'' is the reverse process, when a meteoroid can get captured with low velocity by another planetary system.  The fact that the escape velocities of the meteoroids we consider here are small significantly enhances the probability that a meteoroid can be weakly captured by another planetary system, due to the lower approach velocity to the neighboring stars.

This paper studies the slow chaotic transfer of solid meteoroids between planetary systems within a star cluster, focussing on the transfer probabilities which are a critical factor in the assessment of the possibility of lithopanspermia. We consider the observed size distributions of asteroids and Kuiper belt objects as well as velocity distributions of solar system ejecta from dynamical simulations, to estimate the number of very low velocity ejecta, and thereby estimate the number of weak transfer events in the solar birth cluster. Section \ref{sec:model_weaktranfer} describes the model for minimal energy transfer of meteoroids between two stars in the cluster, where the transfer takes places between two ``chaotic layers" around each star; these chaotic layers are created by the gravitational perturbations from the most massive planet in the planetary system and from the rest of the cluster stars. We refer to this transfer mechanism as ``weak transfer". Section \ref{sec:orderofmag} describes the location of these chaotic layers and uses geometrical considerations to obtain an order-of-magnitude estimate of the probability that meteoroids that have weakly escaped a star are weakly captured by its nearest neighbor in the cluster.  The latter calculation is refined in Section \ref{sec:mc} using Monte Carlo simulations. Section \ref{sec:transf_ss} calculates the number of weak transfer events between the young solar system and the nearest star in the cluster; we explore two cases, where the target star is a solar-type and a low-mass star.  Finally, in Section \ref{sec:conclusions} we summarize our results and discuss the implications for lithopanspermia.

\section{Minimal energy transfer of solid material within a star cluster}
\label{sec:model_weaktranfer}

In this section, we first introduce the concepts of weak capture and weak escape (see Belbruno 2004, 2007, 2010 for a detailed discussion) and then describe a model of how to construct a minimal energy transfer of solid material within a star cluster via weak transfer. 

\subsection{General planetary system model}

Consider a general planetary system ($S$) consisting of a central star ($P_1$) and a system of $N$ planets ($P_i, i=2,...,N$, with $N \geq 3$) on co-planar orbits that are approximately circular, where the labeling is not reflective of the relative distances from $P_1$. Assume that the mass of the star ($m_1$) is much larger than the masses of any of the planets ($m_i$, i.e. $m_1 \gg m_i$,  for  $i=2,...,N$) and that the mass of one of the planets ($P_2$) is much larger than the masses of all the other planets (i.e. $m_2 \gg ~m_i$, for $i = 3,...,N$ -- this condition is fulfilled in the case of our solar system where $P_2$ is Jupiter). A meteoroid ($P_0$) is considered to have a negligible mass ($m_0 = 0$) with respect to the mass of any of the planets and therefore does not gravitationally perturb the circular orbit of $P_2$. Without loss of generality, we view $S$ to consist of $P_1$ and the planet $P_2$, moving around $P_1$ in a circular orbit of radius $\Delta <$ 500 AU. The motion of $P_0$ within this system is  the classical three-dimensional restricted three-body problem, hereafter $RP3D$.  If $P_0$ is constrained to move in the plane of motion of $P_2$ and $P_1$, we have the planar circular restricted three-body problem, hereafter $RP2D$. The differential equations for $RP3D$ are well known (see Belbruno 2004).

Because we are interested in the possibility of lithopanspermia to and from the solar system, in this paper we have assumed that $P_2$ is a Jupiter-mass planet.  However, recent observational results by the Kepler mission indicate that planetary systems with several Neptune-mass planets maybe more common (Batalha et al., in prep.).  The weak transfer mechanism described in this paper could also be applied in this latter case and in some regards enhanced. We leave the study of these cases for future work. 

\subsection{Weak capture and weak escape} 

A convenient way to define the capture of $P_0$ with respect to $P_1$ or $P_2$ in $RP3D$ is by using the concept of ``ballistic capture''. The two-body Kepler energy ($E_k$) of $P_0$ with respect to one of the bodies $P_k$ ($k=1,2$) is

\begin{equation}
E_k = {1\over2}v_k^2 - {m_k\over r_k}, 
\end{equation}												

\noindent where $v_k$ is the velocity magnitude of $P_0$ relative to $P_k$ and $r_k$ is the distance of $P_0$ from $P_k$. The Kepler energy is a function of the trajectory, $E_k({\bf \mbox{\boldmath$\phi$}}(t))$, where ${\bf \mbox{\boldmath$\phi$}}(t) = ({\bf r}(t), {\bf v}(t))$ is the solution of $RP3D$ for the trajectory of $P_0$ in inertial coordinates,  $t$ is time, ${\bf r} = (r_1, r_2, r_3)$ is the position vector from the center of the inertial coordinate system to $P_0$ and ${\bf v} = (v_1, v_2, v_3)$ is the velocity vector of $P_0$.  Ballistic capture takes place when $E_k < 0$, while ballistic escape occurs at the transition from $E_k = 0 $ and $E_k > 0$.  
We are interested when $P_0$ changes from hyperbolic motion with respect to $P_2$ to ballistic capture. 
These trajectories are referred to as ''weak capture'' and  Belbruno (2004, 2010) showed they  are generally unstable and chaotic. The region around $P_2$ where weak capture occurs in position-velocity space is called a weak stability boundary (WSB\footnote{A detailed mathematical explanation of the WSB can be found in Belbruno et al (2010) and Garc\'{\i}a \& G\'omez (2007). They showed that the weak capture boundary is a complicated region that has a fractional dimension and that is equivalent to a Cantor set.  The WSB region can be viewed as a limit set of the stable manifolds to the Lyapunov orbits associated to the collinear Lagrange points $L_1, L_2$ of $P_2$.}). This region around $P_2$ ($WSB(P_2)$) results from the gravitational perturbation of $P_1$. It can be viewed as a location where a particle is tenuously and temporarily captured by $P_2$: the particle will first move around $P_2$ for a short time with negative Kepler energy that approaches zero and then increasing above zero yielding to hyperbolic escape. Ballistic escape from $P_2$ is referred to as ``weak escape" because it occurs near $WSB(P_2)$.      

As is evident by numerical integration of $P_0$ around $P_2$, weak capture generally occurs for relatively short time spans. For example, in the case of the Earth-Moon system (where $\mu = \frac{m_2}{m_1+m_2} = 0.012$), weak capture around the Moon occurs for time spans of days or weeks; in the case of the Sun-Jupiter system, the timescale would be months to a few years (Belbruno \& Marsden, 1997; Belbruno, 2007). 

In our solar system, the existence of the weak stability boundary and the viability of weak capture was demonstrated by the Japanese spacecraft {\it Hiten}: using a trajectory designed by Belbruno (1993, 2007), {\it Hiten} was captured into an orbit around the Moon in 1991 without the use of rockets to slow down.  Weak capture at the Moon was also achieved in 2004 by the ESA spacecraft {\it SMART1} (Racca, 2003; Belbruno, 2007). In another application, weak escape from the Earth's L$_4$~(or L$_5$) Lagrange point was invoked to suggest a low energy transfer to the Earth for the hypothetical Mars-sized impactor that is thought to have triggered the ``giant impact'' origin of the Moon (Belbruno \& Gott,  2005). 

\subsection{Parabolic motion and low velocity escape from $S$ } 

In the absence of $P_2$ in the $RP2D$, a parabolic orbit $Q(t)$ for $P_0$ around $P_1$ is a planar two-body Keplerian parabolic trajectory.  When $P_2$ is considered, there is a ``chaotic layer'' in  $Q$-space near the parabolic trajectory that consists of infinitely many parabolic and near-parabolic trajectories (Xia, 1992; Belbruno, 2004).  This layer has a positive measure and a two-dimensional transversal slice  in four-dimensional position and velocity space yields a fractal structure.  This layer is obtained when a parabolic trajectory is approximately at its periapsis with respect to $P_1$, and where it also has an approximate periapsis passage with respect to $P_2$, slightly beyond $P_2$ (i.e. at a radial distance from $P_1$ slightly larger than $\Delta$).  The periapsis passage near $P_2$ is done in such a way so that $P_0$ is slightly hyperbolic with respect to $P_2$ (approximately in $WSB(P_2)$).  This imparts a small gravity assist to $P_0$. In this case, there are infinitely many possible trajectories near the original parabolic trajectory, some which do not move out infinitely far from $P_1$ and eventually fall back towards $P_1$ for another possible flyby of $P_2$. Other near-parabolic trajectories will escape $P_1$ on hyperbolic trajectories with respect to $P_1$ with a positive escape velocity, $ \sigma$. Because these hyperbolic trajectories lie near parabolic trajectories, $\sigma$ will be small. We  refer to these as ``low velocity escape trajectories" or ``minimal energy escape trajectories". Within the chaotic layer mentioned above exist infinitely many parabolic trajectories that hyperbolically escape $P_1$ with small escape velocity $\sigma$. The distance $R = R_{\rm esc}(m_1)$  from $P_1$ at which $P_0$ achieves this escape velocity can be obtained by the formula $\sigma = \sqrt{\frac{2Gm_1}{R}}$, where $G$ is the gravitational constant.  The fact that these escape trajectories are nearly parabolic and have very small escape velocities implies that their Kepler energy with respect to $P_1$ is nearly zero, for sufficiently large $R$. 

In summary, in the circular restricted three-body problem ($P_0$, $P_1$, $P_2$), there is a chaotic layer that replaces the regular parabolic trajectories of the two-body problem ($P_0$, $P_1$); weak transfer can take place in this chaotic layer because this set of trajectories have low escape velocities. 

\subsection{Model for the minimal energy transfer of solid material within a cluster}

To achieve a minimal energy transfer of the meteoroid $P_0$ from $S$ to another system $S^*$, consider that the system $S$ is  embedded in a star cluster; this will introduce additional gravitational perturbations on $P_0$ which will interact with the gravitational field of $P_1$ and form a weak stability boundary around $P_1$ where the motion of $P_0$ will be chaotic. Because the transfer mechanism requires low relative velocities, we consider an open star cluster with a low velocity dispersion, with relative stellar velocities  $U$ $\approx$ 1 km/s (for comparison, older globular clusters have stellar velocity dispersions of several tens of km/s - Binney \& Tremaine 1988, Meylan \& Heggie 1997).  After the cluster disperses, the relative distances and relative velocities between a star and its closet neighbors increase significantly. For example,  the Sun's current closest neighbor $\alpha$-Centauri (not necessarily a cluster sibling) is now $2.6\cdot10^5$ AU away  (4.28 pc) moving with a relative velocity of 6 km/s. The latter is significantly higher than the $\sim 1$~km/s required for the weak transfer mechanism. In this paper we consider the transfer of solid material between stars in a cluster before the cluster starts to disperse. 

Imagine that $P_0$ makes a minimal energy escape from $P_1$ in the $P_1,P_2$ orbital plane by making a slightly hyperbolic flyby of $P_2$ at $WSB(P_2)$ (where the latter is the weak stability boundary around $P_2$ created by the gravitational perturbation of $P_1$);  $P_0$ moves away from $P_1$ with a escape velocity $\sigma$ near zero. 
$P_0$ then gets to $WSB(P_1)$, the weak stability boundary of $P_1$ caused by the gravitational perturbation of the other $N-1$ stars in the cluster (where $CS_{\rm N-1}$ represents this set of N-1 stars, $\{P_2, P_3, ... , P_N\}$); this boundary is located at a distance $R_{\rm esc}(m_1)$ from $P_1$, at which the combined gravitational force of the stars in the cluster is comparable to the gravitational force of $P_1$. As described previously, the motion in this region is chaotic and lies in the transition between capture and escape from $P_1$\footnote{The sensitivity of the motion of $P_0$ at the distance $R$ can be deduced from an analogous four-body problem described in Belbruno (2004) and Marsden \& Ross (2006) for a transfer to the Moon used by the spacecraft {\it Hiten}. In this case, the four bodies are the Earth ($P_1$), Moon ($P_2$), spacecraft ($P_0$), and the Sun, analogous to $CS_{\rm N-1}$.  The spacecraft leaves the Earth, flys by the Moon, where the flyby is weakly hyperbolic, and then travels out to roughly 1.5$\cdot$10$^6$ km where the gravitational force of the Sun acting on the spacecraft approximately balances that of the Earth. At this location, the motion of the spacecraft is highly sensitive to small differences in velocity and lies between capture and escape from the Earth, i.e. lies at the weak stability boundary between the Earth and Sun.  The motion of $P_0$ in this region is chaotic and  the path of $P_0$ can be altered using a very small amount of fuel.}. Although $WSB(P_1)$ is a complicated fractal set, non-spherical in shape, the sphere of radius $R_{\rm esc}(m_1)$ around the central star lies approximately in the part of $WSB(P_1)$ where the motion is least stable and the trajectories are slightly hyperbolic with respect to $P_1$ (Belbruno et al. 2010; Garc\'{i}a \& G\'omez, 2007) . We refer to trajectories that escape $P_1$ from $WSB(P_1)$ as ``weak escape" trajectories (note that low-velocity escape does not imply weak escape unless it occurs at a weak stability boundary).  The structure of the set of near-parabolic trajectories that escape $P_1$ by flyby of $P_2$  yield infinitely many such trajectories. In particular, this implies that for any point $p$ on the circle of radius $R_{\rm esc}(m_1)$ around $P_1$, there is a nearly parabolic trajectory that will pass by $p$ with velocity $\sigma$, weakly escaping $P_1$. This is illustrated in Figure 1 (see also Figure 1 in Moro-Mart\'{\i}n $\&$ Malhotra, 2005). 

We now compute the trajectory of $P_0$ from the distance $WSB(P_1)$ until it encounters another planetary system $S^*$ centered on star $P_1^*$ of mass $m_1^*$. This is done by using a different set of differential equations; during this motion the trajectory of $P_0$ is relatively undisturbed.  The meteoroid is then weakly captured at $WSB(P_1^*)$ at the distance $R^* =  R_{\rm cap}(m_1^*)$ from $P_1^*$. Figure 1 gives a schematic representation of this process. After $P_0$ is weakly captured at the distance  $R_{\rm cap}(m_1^*)$ from $P_1^*$, say at time $t=T$, it moves towards $P_1^*$ for $t > T$.  Analogous to system $S$, we assume that system $S^*$ has a dominant planet $P_2^*$ orbiting $P_1^*$  in a circular orbit at a radial distance $\Delta^*$. As $t$ increases from $T$, $P_0$ will move to a periapsis distance $r_p^*$ from $P_1^*$, where $r_p^*$ can range from 0 (collision with $P_1^*$) to $R_{\rm cap}(m_1^*)$. However, unlike the escape of $P_0$ from $P_1$, where $P_0$ had a periapsis near the location of $P_2$ and in the same orbital plane of $P_2$, the approach of $P_0$ into the $S^*$ system is three-dimensional and need not be restricted to lie in the $P_2^*$ orbital plane, nor have a periapsis of $P_1^*$ near $P_2^*$. The resulting motion of $P_0$ for $t > T$ will, in general, be complicated and a priori not known. There is no way to predict if $P_0$  will pass near or collide with $P_2^*$ without doing numerical simulations that are outside the scope of this paper. For this reason, we conclude our search at $t=T$, when $P_0$ is weakly captured at $P_1^*$, and calculate the probability of this weak capture to take place. An order-of-magnitude estimate of the weak capture probability is derived in Section \ref{sec:orderofmag} based on geometrical considerations, while in Section \ref{sec:mc} we calculate the probability numerically using Monte Carlo simulations. 


\section{Order of magnitude estimate of the weak capture probability}
\label{sec:orderofmag}

\subsection{Approximate location of the weak stability boundary}	
\label{subsec:wsb}

We calculate $R_{\rm esc}(m_1)$ and $R_{\rm cap}(m_1^*)$ as a function of the stellar mass. For weak escape to take place, the velocity $\sigma$ of the meteoroid at the distance $R_{\rm esc}(m_1)$ from the star $P_1$ must be sufficiently small. Because we are considering slow transfer within an open cluster with a characteristic dispersion 
velocity $U \approx 1$ km/s, we require that $\sigma$ is significantly smaller than $U$, i.e. of the order of 0.1 km/s. This is much smaller 	than the nominal values of several km/s used by the Monte Carlo methods in Melosh (2003) and Adams \& Spergel (2005). 									

To place the above choice for $\sigma$ in context, we study the velocity distribution of weakly escaping test particles from the solar system using a three-body problem between the Sun ($P_1$), Jupiter ($P_2$) and a massless particle ($P_0$). To be consistent with our framework, we model this using $RP2D$, where the test particle moves in the same plane of motion as Jupiter, assumed to be in a circular orbit at 5 AU. The trajectory of the test particle is numerically integrated by a standard Runge-Kutta scheme of order six and numerical accuracy of $10^{-8}$ in the scaled coordinates. The initial conditions of the test particle is an elliptic trajectory very close to parabolic with periapsis distance $r_p$=5 AU and apoapsis distance $r_a$=40,000 AU.  (Note that such orbits are not dissimilar to those of known long period comets in the solar system.) For each numerical integration, we assume that Jupiter is at a random point in its orbit when the test particle starts from apoapsis at 40,000 AU and falls towards $P_1$. We record the velocity of the test particle at the time it escapes with respect to the Sun (i.e. when the Kepler energy with respect to the Sun is positive) after performing a sufficient number of Jupiter fly-bys. Figure 2 shows the distribution of $v_\infty$: out of 670 cases, 58\% have $v_\infty \leq$ 0.1 km/s and 79\% have $v_\infty \leq$ 0.3 km/s. Based on these results, we will assume the velocity $\sigma$ of the meteoroid at the distance $R_{\rm esc}(m_1)$ from the star to be in the range 0.1--0.3 km/s. 


For a given $\sigma$, the location of the part of the weak stability boundary that gives the most unstable motion is approximately given by the sphere of radius $R_{\rm esc}(m_1) = 2Gm_1/\sigma^2$, where $m_1$ is the mass of the star and $\sigma$ is in the range 0.1--0.3~km/s (see Figure 3). Beyond this boundary, when $P_0$ weakly escapes $P_1$, we assume that the meteoroid will move at a nearly constant velocity $\sigma$ with respect to the star.


Slow transfer to a neighboring planetary system enables the meteoroid to arrive at the distance $R_{\rm cap}(m_1^*)$ with a relative velocity with respect to the target star ($P_1^*$) that is similar or smaller than its parabolic escape velocity at that distance, $\sigma^* = \sqrt{\frac{2Gm^*_1}{R_{\rm cap}(m_1^*)}}$, where $m^*_1$ is the mass of the target star;  if its velocity is higher than $\sigma^*$, it will not be captured and will fly by.  Since the relative velocity between stars in the cluster is $U \approx 1$~km/s, the meteoroid that weakly escaped from star $P_1$ moves toward the target star $P_1^*$ with velocity $U \pm \sigma$ (see Figure 1). Because $\sigma$ is small relative to $U$ and can be neglected, the relative velocity of the meteoroid with respect to the target star is $\approx U$. Therefore, weak capture can occur at the distance at which $\sigma^* \approx U \approx 1$~km/s, i.e. $R_{\rm cap}(m_1^*) = \frac{2Gm_1^*}{U^2}$. Figure 3 shows $R_{\rm esc}(m_1)$ and $R_{\rm cap}(m_1^*)$ as a function of the stellar mass. The horizontal lines indicate the range of cluster sizes (dashed-dotted lines) and mean interstellar distances (dotted lines) for clusters consisting of N = 100, 1000 and 4300 members respectively; clusters with this range of sizes  are the birthplaces of a large fraction of stars in the Galaxy (Lada \& Lada 2003). The radius of the cluster depends on the number of stars and is given by $R_{\rm cluster} = 1 \rm{pc} \sqrt{N/300}$ (Adams 2010). For N=100, 1000 and 4300, we get that $R_{\rm cluster}$ is approximately $2.1\cdot10^5$ AU, $6.5\cdot10^5$ AU and $1.3\cdot10^6$ AU respectively. We can estimate the average interstellar distance within a cluster, $D = n^{-1/3}$, where $n = \frac{3N}{4\pi R_{\rm cluster}^3}$ is the average number density of stars in the cluster.  $D$ is approximately 7$\cdot$10$^4$ AU,  10$^5$ AU and 1.2$\cdot$10$^5$ AU, for a cluster with N=100, 1000 and 4300 members respectively.  The choice of N =4300 is explained in Section \ref{sec:clusprop}.

It is of interest to compare the results in Figure 3 to those in Melosh (2003). The latter considers the exchange of meteoroids between field stars using hyperbolic trajectories; it estimates a cross-section of 0.025 AU$^2$ for meteoroid capture by a planetary system with a Jupiter-mass planet located at 5 AU. Under the scenario considered in the present work, the order-of-magnitude estimate for the weak capture cross-section would be $\pi \cdot R_{\rm cap}^2$; for a solar-mass star, $R_{\rm cap} \sim $ 2$\cdot$10$^3$ AU (Figure 3) so that the weak capture cross-section is $\sim$ 10$^7$ AU$^2$ (i.e. many order of magnitudes larger than Melosh 2003 estimate). This indicates that weak escape and capture within an open cluster can enhance drastically the probability of transfer. A recent study in which enhanced capture probability within a stellar cluster has been invoked is that of Levison et al. (2010). They modeled the exchange of comets within a stellar cluster assuming that each star is surrounded by a large disk of comets with $q$ = 30 AU and $a$ = 1000--5000 AU, and with a population similar to that of the Sun; they conclude that up to 90\% of comets in the Oort Cloud may be have an extrasolar origin. 

\subsection{Constraints on Stellar Masses for Weak Transfer}
\label{subsec:cons}

From Figure 3, we can set constraints on the stellar mass $m_1$ that could allow weak escape from $P_1$ to take place. The idea is simple: if for a given $\sigma$ (within the range 0.1--0.3 km/s, see Section \ref{subsec:wsb}), we have that $R_{\rm esc}(m_1) < D$, i.e.~the weak escape boundary is located within the distance of the next neighboring star $P_1^*$, then weak transfer is possible because at the time the meteoroid passes near the star $P_1^*$, its velocity is similar to the mean stellar velocity dispersion ($U$ $\sim1$~km/s) and there is a significant probability of capture (which we quantify in Section \ref{subsec:capx}). Conversely, weak transfer by the process schematically represented in Figure 1 is much less likely to take place if $R_{\rm esc}(m_1)> D$ because at the time the meteoroid passes near a neighboring star, its velocity is too high and, as a consequence, it will likely fly by.   Figure 3 shows that for $\sigma=0.1$~km/s, the condition $R_{\rm esc}(m_1) <  D$ for weak escape is satisfied for $m_1 <  0.4 M\odot$, $m_1 <  0.57 M\odot$ and $m_1 <  0.68 M\odot$, for clusters with N = 100, 1000 and 4300 members respectively. If we consider the higher but still acceptable value $\sigma=0.3$~km/s, Figure 3 shows that the stellar mass limits for weak escape are $m_1 < 3.5M_{\odot}$,  $m_1 < 5.1 M_{\odot}$ and $m_1 < 6.1M_{\odot}$, for clusters with N =100, 1000 and 4300 members respectively. A meteoroid escaping nearly parabolically from any star with a mass larger than these limits will achieve a  velocity of $\sigma<0.1$--0.3~km/s only at a distance larger than the mean interstellar distance in the cluster; weak transfer is not likely to take place under such conditions. 

Of particular interest is the case of the solar system as the source of the meteoroids. It has been estimated that the Sun's birth cluster consisted of N=4300$\pm2800$ members (Adams 2010).  For a $1 M_{\odot}$ star in such a cluster, we find that the parabolic escape velocity at the mean interstellar distance of D = 1.2$\cdot10^5$ AU is $\sigma = \sqrt{2Gm_1/D} = 0.12$ km/s.  Because this values of $\sigma$ lies within the range of values of interest for weak escape, we conclude that the meteoroids originating in the early solar system could  have met the conditions for weak escape in the Sun's birth cluster.

We have established the range of stellar masses that could in principle allow weak escape to take place. We now estimate the probability of weak capture by a neighboring planetary system. Whether or not the transfer of meteoroids from one star to any star of a given mass $m_1^*$ takes place will depend on the relative capture cross-section of the target star and on the number of potential targets. We discuss these factors below. 

\subsection{Relative capture cross-section of the target star}
\label{subsec:capx}

In the case of weak transfer, assuming that the meteoroids escape isotropically from $P_1$, the relative capture cross-section of the target star would be  given by $ C_S = {G_f}(R_{\rm cap}(m_1^*)/D)^2$, where $D$ is the distance between the two stars, $R_{\rm cap}$ is the weak capture boundary (illustrated in Figure 1 and estimated in Figure 3) and the factor $G_f$ accounts for gravitational focussing. The assumption of isotropy is not justified in the framework of the restricted three-body problem, but it is expected in the presence of the cluster stars and of the galactic potential that can isotropize the escaping meteoroids' orbits, in analogy with the isotropization of the Oort Cloud of comets (Dones et al.~2004).  

The gravitational focussing factor is given by 

\begin{equation}
G_f = 1 + ({v_{\rm esc} \over v_{\infty}})^2,
\end{equation}

\noindent where $v_{\infty}$ is the velocity at infinity and $v_{\rm esc}$ is the escape velocity at the distance $R_{\rm cap}(m_1^*)$ from $P_1^*$. In the case of weak capture, the term $G_f$ increases the cross-section due to enhanced gravitational focussing.  For example, as we saw previously in the case of the Sun and Jupiter, $v_{\infty} \approx 0.1-0.3$ km/s, while at the distance $R_{\rm cap} = 40000$ AU we have that $v_{\rm esc} \approx 0.2$ km/s. In this case we get that $G_f \approx 2$, doubling the capture cross-section. This situation occurs in our capture methodology. In the formulation for determining $R_{\rm cap}(m_1^*)$, the meteoroid has an approximate relative approach velocity to the target star $P_1^*$ of roughly $U = 1$ km/s, which represents $v_{\infty}$. However, $R_{\rm cap}$ is determined so that this same value of velocity is taken for $v_{\rm esc}$ from the target star. That is, we are assuming $v_{\infty} \approx v_{\rm esc} \approx 1$ km/s. This implies also that $G_f = 2$. 
This is a conservative estimate and does not make use of the nature of weak capture dynamics. In fact, the value of $G_f$ can be substantially increased, as we demonstrate below, if at a given value of $R_{\rm cap}(m_1^*)$, $v_{\infty}$ is  smaller than the approximate value of $U = 1$ km/s, while $v_{\rm esc}$ $\approx$ $U$; in this case, $v_{\infty} \rightarrow$ 0 which implies $G_f \rightarrow \infty$. 

\subsection{Number of potential targets}
\label{sec:oomprob}

We denote $P_{\rm IMF}(m_1^*)$ as the probability of finding a star of mass $m_1^*$ in the star cluster. This probability is referred to as the initial mass function (IMF) and it can be inferred from observations of stellar clusters. It is found that a wide range of clusters, varying from large clusters like the Trapezium to smaller clusters like Taurus, as well as older field stars, show very similar distributions of stellar masses down to the hydrogen burning limit at $\sim$ 0.1 $M_{\odot}$ (Lada \& Lada 2003 and references therein).  The IMF can be characterized by the broken power-law, $\xi$(M) = $\xi$$_1$M$^{-2.2}$ for 0.6--100 $M_{\odot}$, and $\xi$(M) = $\xi$$_2$M$^{-1.1}$ for 0.1--0.6 $M_{\odot}$, where $\xi$(M)dM is the number of stars with mass (M, M+dM).  There is a steep decline into the substellar brown dwarf regime and a possible second peak but we will ignore objects below the hydrogen burning limit.  The resulting average stellar mass is $\sim$0.88 $M_{\odot}$. To calculate $P_{\rm IMF}(m_1^*)$  (square symbols in Figure 4), we use a logarithmic binning of masses with d(logM)=0.1 and normalize the distribution to unity, which gives $\xi$$_1$=0.19 and $\xi$$_2$=0.34. 


An upper limit to the probability that a meteoroid escaping from a star of mass $m_1$ (within the range described in Section \ref{subsec:cons}) will get captured by a neighboring star of mass $m_1^*$ is approximately given by the relative capture cross-section of its weak stability boundary, $G_f(R_{\rm cap}(m_1^*)/D_{m_1^*-m_1})^2$, where $R_{\rm cap}$ is the radius of the weak stability boundary for capture of the target star and $D_{m_1^*-m_1}$ is the average distance between two stars of masses $m_1^*$ and $m_1$ respectively. If we assume that this distance is the average interstellar distance $D$ estimated in Section \ref{subsec:wsb}, we get that for a solar-type star, an upper limit to the weak capture probability is approximately 1.3$\cdot 10^{-3}$, 6.5$\cdot 10^{-4}$ and 4.5$\cdot 10^{-4}$, for a cluster of N = 100, 1000 and 4300 members respectively (see dashed line of Figure 4).  However, the distance to a star of a given stellar mass will not be $D$ necessarily, but it will depend on the distribution of stellar masses in the cluster (the IMF). The simplest case is when both stars have equal masses $m_1^*$ = $m_1$.   In this case, the average interstellar distance would be $D_{m_1} \sim (1/n_{m_1})^{1/3}$,  where $n_{m_1}$ is the average number density of stars with mass ${m_1}$,  $n_{m_1} = \frac{3N_{m_1}}{4\pi R_{\rm cluster}^3}$, and $N_{m_1}$ is the total number of stars in the cluster with mass ${m_1}$, $N_{m_1} = N \cdot P_{\rm IMF}(m_1)$.  Because $n_{m_1} = n \cdot P_{\rm IMF}(m_1)$,  we get that $D_{m_1} \sim D \cdot (P_{\rm IMF}(m_1^*))^{-1/3}$. This means that the transfer probability between two stars of equal mass is given by $G_f(R_{\rm cap}(m_1^*)/D)^2 \cdot (P_{\rm IMF}(m_1^*))^{2/3}$,  where $D$ is the average distance between any two stars in the cluster (regardless of their mass). As we have discussed above, the value of the focussing factor $G_f$ can be very large. A conservative estimate of the capture probability can be done by assuming $G_f$ = 2. Figure 4 shows that the capture probability between two planetary systems with solar-type central stars  (m$_1^*$ = m$_1$ = 1 $M_{\odot}$) are 1.9$\cdot 10^{-4}$, 8.1$\cdot 10^{-5}$ and 5.6$\cdot 10^{-5}$, for a cluster of N = 100, 1000 and 4300 members respectively. 

\section{Numerical estimate of the weak capture probability based on Monte Carlo simulations}
\label{sec:mc}

The order-of-magnitude estimate of the weak transfer probability discussed in Section \ref{sec:orderofmag} has two important caveats: (a) it assumes that the capture takes place in the orbital plane of $P_2^*$ (the primary planet in the capturing system), i.e. it does not consider the 3-D problem; and (b) the adoption of a focussing factor $G_f$ = 2 is conservative because it does not make use of the nature of nonlinear weak capture dynamics, where the gravitational forces of $P_1$, $P_1^*$, and the stars of the cluster are all acting on $P_0$.  $G_f$ can increase very significantly if at a given value of $R_{\rm cap}(m_1^*)$, $v_{\infty}$ is  much smaller than the approximate value of $U = 1$ km/s, and can become 0, while $v_{\rm esc}$ $\approx$ $U$ (for an example, see Appendix A). In this section, we refine the estimate of the weak transfer probability using Monte Carlo simulations that address the caveats mentioned above by considering the more general and realistic model that calculates the motion of $P_0$, $P_1$, $P_1^*$ by a general three-dimensional Newtonian three-body problem {\it plus} the effective gravitational perturbation of the cluster stars.  We sample millions of trajectories for the meteoroid $P_0$ weakly escaping $P_1$ testing whether  they are weakly captured near $P_1^*$.  

\subsection{Modeling procedure}

To model the motion of $P_0$ from the distance $R_{\rm esc}(m_1)$ from $P_1$ to $S^*$ and with initial velocity $\sigma$, we consider a general Newtonian three-dimensional three-body problem that is perturbed by the gravitational force of the stellar cluster. This model gives the motion of $P_0, P_1, P_1^*$ in an inertial coordinate system $(r_1, r_2, r_3)$. We assume the gravitational perturbing force of the cluster is obtained from a spherically symmetric Hernquist potential

\begin{equation}
U_C (r)= {{GM}\over{r+a}},
\label{eq:Hernquist}
\end{equation}

\noindent  (Hernquist 1990), where $M$ is the total cluster mass, $r^2 = r_1^2 + r_2^2 + r_3^2$, and $a$ is the cluster scale length, which is approximately the radius of the cluster, $R_{\rm cluster}$.  The center of mass of the cluster is at the origin, ${\bf r}=(0,0,0)$.

The system of differential equations modeling the motion of $P_0, P_1, P_1^*$, of mass $m_0, m_1, m_1^*$ respectively, is given by
\begin{eqnarray}
{\bf \dot{r}}_k  =  {\bf v}_k , \hspace{1cm}
{\bf \dot{v}}_k  =  m_k^{-1} \frac{\partial V}{\partial {\bf r}_k},
\label{eq:DE}
\end{eqnarray}

\noindent where ${\bf r}_k = (r_{k1}, r_{k2}, r_{k3})$ and ${\bf v}_k = (v_{k1}, v_{k2}, v_{k3}) = (\dot{r}_{k1},\dot{r}_{k2}, \dot{r}_{k3})$ are the position and velocity vectors, respectively, of the $k$th particle, $P_k$ (with $k$ = 0, 1, 2;  $P_2$ represents $P_1^*$ and $m_2$ represents $m_1^*$). The gravitational potential is  $V = U + m_kU_C(r_k)$, where $r_k$ is the distance of $P_k$ to the origin, $r_k = |{\bf r}_k|$, and   

\begin{equation}
U = \sum_{j=0 \atop j \neq k} ^2 \frac{G m_j m_k}{r_{j/k}}. 
\end{equation}

\noindent $U = U({\bf r}_1,{\bf r}_2, {\bf r}_3)$ is a function of $9$ variables $r_{kj}, j=1,2,3$, and

\begin{equation}
\frac{\partial V}{\partial {\bf r}_k} \equiv
\left( \frac{\partial V}{\partial r_{k1}},
\frac{\partial V}{\partial r_{k2}},
\frac{\partial V}{\partial r_{k3}} \right), 
\end{equation}

\noindent where $r _{j/k}$ is the distance between $P_j$ and $P_k$,  $r _{j/k} = |{\bf r}_j - {\bf r}_k|$. 	 

\noindent $m_0$ is approximately $0$ relative to $m_1, m_1^*$. 

In the Monte Carlo model, the trajectory of motion for $P_0$ is determined from (\ref{eq:DE}) by providing its initial position at the distance $R = R_{\rm esc}(m_1)$ from $P_1$ at $t=0$. The initial velocity of $P_0$ with respect to $P_1$ at the distance $R$ is chosen from a distribution of weak escape velocities (see Figure 2).  $P_0$ then weakly escapes $P_1$.    
The initial positions and velocities of $P_1$ and $P_1^*$ are also given at $t=0$. Their initial separation distance depends on the cluster properties and is a function of the number of stars in the cluster N. After the initial positions and velocities of $P_0, P_1, P_1^*$ are given, the trajectory of $P_0$ is propagated for $t>0$. We search for the condition of the first weak capture of $P_0$ with respect to $P_1^*$, given by a negative value of the Kepler energy of $P_0$ with respect to $P_1^*$, $E_1^* < 0$. This implies that  for $P_0$, $v_{\infty} \rightarrow$ 0 , and therefore $G_f   \rightarrow \infty$. This occurs at the weak stability boundary of $P_1^*$, $WSB(P_1^*)$, at a distance $R^* = R_{\rm cap}(m_1^*)$ from $P_1^*$. This stability boundary is formed around $P_1^*$ due to the resultant gravitational perturbation of the $N-1$ remaining stars of the cluster. We are interested in cases where the time of propagation $T$ is smaller than the cluster dispersal timescale, which is a function of the number of stars in the cluster, N. Under the conditions described above and using a Monte Carlo approach, we sample millions of trajectories of $P_0$ weakly escaping $P_1$ using Runge-Kutta-Nystr\"om 12th/10th order, variable step, symplectic integrator (Dormand et al. 1987). The Monte
Carlo method calculates the number of these particles to be weakly captured by $P_1^*$, i.e. the probability of weak transfer. 

\subsection{Cluster properties}
\label{sec:clusprop}

We adopt those cluster properties thought to be representative of the Sun's birth cluster. These properties are inferred from a wide range of physical considerations, including the effect of supernova explosions on the enrichment of short-lived radioactive isotopes in the solar nebula, protoplanetary disk truncation due to photoevaporation from nearby hot stars, and the dynamical disruption of planetary orbits due to close encounters with cluster stars. The observed solar system properties that depend on the Solar birth environment (such as the evidence of short-lived radio-nuclides in meteorites and the dynamical properties of the outer solar system planets and Kuiper belt) led Adams (2010) to conclude that the Sun was born in a moderately large cluster with N = 1000--10000 and $\langle N\rangle$ = 4300 $\pm$ 2800 members, similar to the Trapezium cluster in Orion. For reference, Adams (2010) notes that approximately 50\% of stars are born in systems with N $\gtrsim$ 1000, but for $\sim$ 80\% of these stars the clusters dissolve quickly after 10 Myr; only $\sim$ 10\% of the total number of stars would be born in open clusters with lifetimes of the order of 100--500 Myr. 

Using the initial mass function in Section \ref{sec:oomprob}, we estimate that the total cluster mass is $M_{\rm cluster} = (3784\pm2500)M_{\odot}$. Following Adams (2010), the properties of such a cluster would be as follows. The scale length of the cluster is approximately its size, given by $a = R_{\rm cluster}= 1 \rm{pc} \sqrt{N/300} = 3.78\pm1.5$ pc.  With the stellar number density, $n = 3N/(4\pi R_{\rm cluster}^3)$ = 19 pc$^{-3}$, the average interstellar distance is $D = n^{-1/3}= 0.375$ pc. The typical expected distance of the Sun to the center of mass of the cluster would be $d_{\rm cm}= 2R_{\rm cluster}$/3 = 2.52$\pm1$ pc (using a radial profile for stellar density consistent with a Hernquist potential). The cluster lifetime is approximately $T = 2.3 (M_{\rm cluster}/M_\odot)^{0.6}$ Myr = (135--$535)$ Myr, for N = 1000--10000.  

\subsection{Modeling assumptions}

For the Monte Carlo simulations we adopt the average values of the solar system birth cluster: $N=4300, D=0.375$ pc, $a=3.78$ pc, $d_{\rm cm}=2.52$ pc, and a cluster lifetime $T=322.5$ Myr.  We also make the following assumptions: 
\begin{itemize}

\item For the duration of the simulation, the cluster size and therefore the average distance between the stars, $D$, are kept constant (i.e. we making the simplifying assumption that the cluster disperses instantly at the end of the simulation, at time equal to the average cluster age). 

\item At $t=0$, the initial separation of $P_1,P_1^\star$ is $D$  (coinciding with the average separation of the cluster stars); the center of mass of $P_1 , P_1^\star$, placed on the $Q_1,Q_2$ plane for convenience, is assumed to be a distance $d_{\rm cm}$ from the inertial frame origin, $Q = 0 \equiv (0,0,0)$.  

\item Let $\bf{v_1}$ and $\bf{v_1^*}$ be the initial velocity vectors of $P_1$ and $P_1^*$ respectively (relative to $Q=0$). At $t = 0$, these two vectors lie on the  plane $\sigma_{v0}$ and we assume $|{\bf{v_1}}- {\bf{v_1^*}}|=1$ km/s, $|{\bf{v_1}}|$ $\leq$ 2 km/s and  $|{\bf{v_1^*}}|$ $\leq$ 2 km/s, with the initial angle $\theta$  between $\bf{v_1}, \bf{v_1^*}$ varying over $[0,2\pi]$ on $\sigma_{v0}$.

\item At $t = 0$, $P_0$ escapes from $P_1$ in three-dimensions with the tangential velocity vector $\bf{v_{0/1}}$ (relative to $P_1$); the  magnitude of  $\bf{v_{0/1}}$  is chosen from the weak escape $v_{\infty}$ distribution ($D_{v_{\infty}}$) in Figure 2; note that this figure assumed $P_1$ has a mass $m_1 = 1M_{\odot}$ and a Jupiter size planet orbiting at 5 AU, but we also assume this distribution for the case $m_1= 0.5M_{\odot}$ considered below. At $t = 0$,  when  $P_0$ escapes $P_1$, the distance between the two is  $r_{0/1} = |\bf{r_{0/1}}|$ $\in [0.02, 0.2]$ pc. Let $\phi_0, \theta_0$ be the spherical angles that specify the position of the vector $\bf{r_{0/1}}$ of $P_0$ with respect $P_1$ (note that $\theta_0$ is distinct from $\theta$ and should not be confused with it).  We assume $\phi_0 \in [0, 2\pi]$ and $\theta_0 \in [0, \pi]$, where $\phi_0$ is uniformly distributed in $[0, 2\pi]$,  and $\theta_0$ is sinusoidally distributed over $[0, \pi]$. For $t > 0$, as $P_0$ escapes $P_1$ in three-dimensions, it can approach $P_1^*$ from any direction.

\item We search for weak capture with respect to $P_1^*$ in the time interval $t \in [0, T_{\rm max}]$Myr, where $T_{\rm max}$ is the estimated age of the cluster $T_{\rm max} = T = 322.5$ Myr. The condition for capture is a negative value of the two-body Kepler energy of $P_0$ with respect to $P_1^*$, $E_1^* < 0$.  This can occur at any distance from $P_1^*$, or even at the initial  time $t=0$ when $P_0$ escapes $P_1$ (the latter can occur  because $P_0$ escapes $P_1$ between 0.02 pc and 0.2 pc, but $P_1^*$ is only initially at a distance $D$ from $P_1$).  When weak capture occurs, we have obtained a weak transfer of $P_0$ from $P_1$ to $P_1^*$. 
\end{itemize}

\noindent
In summary, the Monte Carlo simulation is done by randomly choosing values of the paramaters $v_{0/1}, r_{0/1}, \theta_0, \phi_0, 
\theta, |\bf{v_1}|$ and $|\bf{v_1^*}|$ at $t=0$ that satisfy: $v_{0/1} \in D_{v_{\infty}}$, $r_{0/1} \in [0.02, 0.2]$ pc,  $\phi_0 \in [0, 2\pi]$,  
$\theta_0 \in [0, \pi]$, $\theta \in [0, 2\pi]$, $|{\bf{v_1}}| \leq 2$ km/s,  $|{\bf{v_1^*}}| \leq 2$ km/s and $\vert \mathbf{v_1} - \mathbf{v_1^\star}\vert = 1$ km/s respectively. The Monte Carlo algorithm searches 
for the condition $E_1^*<0$ for $t \in [0, T_{\rm max}]$ Myr.    

\subsection{Results from the Monte Carlo simulations}
\label{sec:mc_results}
\noindent
The Monte Carlo simulations study the capture probability within the cluster described in Section \ref{sec:clusprop} for the following three different combinations of stellar masses: $m_1 = 1$ M$_\odot$, $m_1^* = 1$ M$_\odot$ (Case 1); $m_1 = 1$ M$_\odot$, $m_1^* = 0.5$ M$_\odot$ (Case 2); and $m_1 = 0.5$ M$_\odot$, $m_1^* = 1$ M$_\odot$ (Case 3).  These simulations explore the weak transfer of material between two solar-type stars and between a solar-type star and a star half its mass, the interest of the latter being that low-mass stars are the most abundant in the galaxy, with $\sim$ 64\% and 72\% having mass $<$ 0.4 M$_\odot$ and 0.6 M$_\odot$ respectively.  
For each case, the Monte Carlo simulation ran 5 million trajectories, a sufficient number so that the randomization of the initial values produces distributions of the parameters that span their respective ranges. The resulting distributions are discussed in the Appendix B. 

For Case 1 (weak transfer of meteoroids between two solar-type stars), the  Monte Carlo simulations give a revised estimate of the weak capture probability of 1.5$\cdot$10$^{-3}$ (see Table 1). Note that the Monte Carlo simulations assume that $P_1^*$ with $m_1^*$ = 1 M$_\odot$ is at the average interstellar distance $D = n^{-1/3}$ (i.e. without regard to the distribution of stellar masses in the cluster). The Monte Carlo derived weak capture probability is approximately a factor of 3 larger than the order-of magnitude-estimate in Section \ref{sec:oomprob}, which was approximated by the relative capture cross-section of weak stability boundary of the target star, given by $G_f(R_{\rm cap}(m_1^*)/D)^2$ = 4.5$\cdot 10^{-4}$, for  $m_1^*$ = 1, N = 4300 and using a conservative gravitational focussing factor of $G_f$ = 2. This increase in the weak capture probability estimated by the Monte Carlo simulations is likely due to the larger focussing factors that result from the nature of weak capture. Note that both estimates assume that the meteoroids are initially on weakly escaping trajectories from $P_1$. For Cases 2 and 3, the Monte Carlo simulations give a weak capture probability of 0.5$\cdot$10$^{-3}$ and 1.2$\cdot$10$^{-3}$ respectively (see Table 1).

\section{Estimate of the number of weak transfer events for the young solar system}
\label{sec:transf_ss}

With the improved estimate of the probability of weak capture obtained from the Monte Carlo simulations, we now calculate the total number of meteoroids that could get transferred between two neighboring solar-type stars harboring planetary systems by multiplying the capture probability in Table 1 (for Case 1) with the estimated number of meteoroids (N$_{\rm R}$) that a planetary system may eject before the cluster disperses. Because the scenario considered here assumes that the escaping meteoroids are on weakly escaping trajectories, N$_{\rm R}$ is not the total number of escaping meteoroids but only the subset on weakly escaping trajectories.  The main uncertainties in assessing whether weak transfer is a viable method for meteoroid transfer lie in estimating this number. While this number is highly uncertain, we proceed with estimating N$_{\rm R}$ for our solar system, as it is the only planetary system for which observations and dynamical models enable us to make an educated estimate. 

We estimate N$_{\rm R}$ from Oort cloud formation models because Oort cloud comets are weakly bound to the solar system and are therefore representative of a population of meteoroids that may have been delivered on its weak stability boundary, subject to weak escape. We use the model by Brasser et al. (2011) in which $\sim$ 1\% of the planetesimals in the Jupiter--Saturn region (4--12 AU) became part of the primordial Oort Cloud in the early history of the solar system. Under this scenario, the forming giant planets scattered the planetesimals in this region out to large distances where they were subject to the slowly changing gravitational potential of the cluster; the latter caused the perihelion distances of the scattered planetesimals to be lifted to distances $\gg 10$ AU, where the planetesimals were no longer subject to further scattering events but were also safe from complete ejection, and thus remained weakly bound to the solar system. To calculate how many planetesimals formed in the 4-12 AU region, we adopt the minimum mass solar nebula (MMSN), with a dust surface density $\Sigma=\Sigma_0(a/40 AU)^{-3/2}$, where $\Sigma_0=\Sigma_{0d}$ = 0.1 g/cm$^2$ (Weidenschilling 1977; Hayashi 1981). Integrating between 4 and 12 AU, we find a total mass in solids of $10^{29}$ g. With the above estimate for the total mass in solids, we calculate the number of planetesimals by adopting a planetesimal size distribution function representative of the early solar system. This size distribution is uncertain, but can be constrained roughly from observations and coagulation models. Based on these studies, we adopt three power-law size distribution functions for our calculation of N$_{\rm R}$, with dN/dD $\propto$ D$^{-q_1}$ for $D>D_0$ and dN/dD $\propto$ D$^{-q_2}$ if D$<$D$_0$, of different power-law indices at the small and large sizes:

\begin{itemize}
\item {\it Case A}: $q_1$ = 4.3, $q_2$ = 3.5, $D_0$ = 100 km, 
D$_{\rm max}$ = 2000 km ($\sim$ Pluto's size),
D$_{\rm min}$ = 1 $\mu$m ($\sim$ dust blow-out size).
This size distribution has the power law index of a collisional cascade at the small size end, and that of the hot Kuiper Belt at the large size end (from Bernstein et al. 2004), with a break diameter consistent with that of the present-day Kuiper belt.  The approximate scenario of this case is as follows: in the Jupiter-Saturn zone, the accretion of large planetesimals proceeded to make the large-size bodies similar to those observed in the present-day Kuiper belt, whereas at the small size end, the dynamical stirring by the large bodies produced a classical collisional cascade. For the latter, the reason why we do not adopt the observed present-day power law index of the hot Kuiper belt at sizes $\lesssim$ 50 km ($q_2$ = 2.8, Bernstein et al. 2004) is because these smaller bodies are the result of an advanced erosional process that changed their size distribution on gigayear timescales (Pan \& Sari 2005). 

\item {\it Case B}: $q_1$ = 3.3, $q_2$ = 3.5, $D_0$ = 2 km, D$_{\rm max}$ = 2000 km ($\sim$ Pluto's size), D$_{\rm min}$ = 1 $\mu$m ($\sim$ dust blow-out size). This size distribution is derived from theoretical coagulation models (see review in Kenyon et al. 2008). 

\item {\it Case C}: $q_1$ = 2.7, $q_2$ = 3.5, $D_0$ = 2 km, D$_{\rm max}$ = 2000 km ($\sim$ Pluto's size), D$_{\rm min}$ = 1 $\mu$m ($\sim$ dust blow-out size). This size distribution is also based on theoretical coagulation models (see review in Kenyon et al. 2008). 

\item {\it Case D}: $q_1$ = 4.3, $q_2$ = 1.1, $D_0$ = 100 km, D$_{\rm max}$ = 2000 km ($\sim$ Pluto's size), D$_{\rm min}$ = 1 $\mu$m ($\sim$ dust blow-out size). This size distribution represents perhaps the ``worst case" scenario in which the depletion of the small bodies took place very early on, before the objects were transferred into the Oort Cloud.  The parameters here are within the range that Bernstein et al.~(2004) find for the Kuiper belt. This size distribution is also similar to that discussed in models for the primordial asteroid belt ($q_1 = 4.5$, $q_2$ = 1.2, $D_0$ $\sim$ 100 km, Bottke et al.~2005). 

\end{itemize}

Of particular interest are the meteoroids $>$10 kg, that may be large enough to shield potential biological material from the hazards of radiation in deep space and from the impact on the surface of a terrestrial planet (Horneck 1993; Nicholson et al. 2000; Benardini et al. 2003; Melosh 2003). 

Given a total mass of $10^{29}$ g of solids in the 4--12 AU zone in the solar system at the time of Jupiter and Saturn formation, and adopting the above size distributions of the solid bodies, we can calculate the number of planetesimals with masses $>10$ kg (diameter $D > 26$ cm if $\rho$ = 1 g/cm$^{-3}$).  Then, assuming that 1\% of these planetesimals were delivered to the weak stability boundary of the solar system, we obtain an estimate for N$_{\rm R}$. Table 2 shows these results for the four size distributions considered. We find that N$_{\rm R}$ is in the range $8 \cdot 10^{16}$--$ 2 \cdot 10^{19}$ for Cases A, B and C, but is only $5 \cdot 10^{6}$ for Case D. 

Finally, multiplying N$_{\rm R}$ by the weak capture probability determined in Section \ref{sec:mc} (1.5$\cdot 10^{-3}$ for the transfer between two solar-type stars), we estimate the number of weak transfer events, $N_{\rm WTE}$, that could have occurred between the early solar system and a neighboring star in the cluster, assuming it is also a solar-type star and harbors a planetary system. The results are listed in Table 2; for Cases A, B and C they are of the order of $10^{14}$--$ 3 \cdot 10^{16}$, and for Case D it is of the order of $10^{4}$.  If the target system is a low-mass star of $m_1^*$ = 0.5 M$_{\odot}$, the number of weak transfer events is approximately three times smaller in each case (because the weak capture probability in this case is 5$\cdot 10^{-4}$ instead of 1.5$\cdot 10^{-3}$ -- see Table 1). 

Note that the results in Table 2 refer to the weak transfer of solid material between neighboring planetary systems, but it does not account for the probability of landing on a terrestrial planet in the target system. Melosh (2003) and Adams \& Spergel (2005) estimate that the latter probability is $\sim 10^{-4}$. 

\section{Summary and Discussion}
\label{sec:conclusions}

\subsection{Summary of the weak transfer mechanism}
\label{sec:summary}
We have explored a mechanism that allows the transfer of solid material between two planetary systems embedded in a cluster. This mechanism is based on the chaotic dynamics of the restricted three-body model of the meteoroid, the star and the most massive planet in the planetary system ($P_0$, $P_1$, $P_2$), in which a chaotic layer replaces the regular parabolic trajectories of the two body problem of ($P_0$, $P_1$). Similarly, there is a chaotic layer around the target star (assumed to harbor a planetary system). Weak transfer takes place within these chaotic layers because the trajectories have low escape velocities and therefore the capture probability in enhanced. We have applied this mechanism to the problem of planetesimal transfer between planetary systems in an open star cluster, where the relative stellar velocities are sufficiently low ($\sim$ 1 km/s) to allow weak escape and capture. We found that weak escape and capture within an open cluster can enhance drastically the probability of transfer compared to the scenario described in Melosh (2003) and corresponding to the exchange of meteoroids between field stars using hyperbolic trajectories: while Melosh (2003) estimated a cross-section of 0.025 AU$^2$ (for capture by a planetary system with a Jupiter-mass planet at 5 AU), and order-of-magnitude estimate for weak capture cross-section would be $\pi \cdot R_{\rm cap}^2 \sim 10^7 AU^2$ (i.e. many order of magnitudes larger). 

To obtain quantitative estimates, we adopt the average cluster properties inferred for the Solar birth cluster (Adams 2010), with N = 4300 members, a total mass of $M_{\rm cluster}$ = 3784 M$_{\odot}$ (using an average stellar mass of 0.88  M$_{\odot}$ resulting from the initial mass function) and a cluster scale length of $a = 1 \rm{pc} \sqrt{N/300}$ = 3.78 pc. Such a cluster is expected to have a lifetime of 2.3 $(M_{\rm cluster}/M_\odot)^{0.6}$ Myr = 322.5 Myr (ranging from 135--535 Myr,  for N = 1000--10000). 

With the aid of Monte Carlo simulations, we estimate the probability of weak capture for meteoroids that have weakly escaped their original planetary system and are weakly captured at a neighboring system. We consider three cases where the source $P_1$ and the target star $P_1^*$ have masses of $m_1$ = 1,   $m_1^*$ = 1 (Case 1), $m_1$ = 1,   $m_1^*$ = 0.5 (Case 2) and $m_1$ = 0.5,   $m_1^*$ = 1 (Case 3). The resulting weak capture probabilities are 0.15\%, 0.05\% and 0.12\% respectively.  This capture probability is much larger than the capture probabilities obtained in previous studies; for example, Adams \& Spergel (2005) found capture probabilities of $\sim10^{-6}$ for mean ejection speeds of $\sim$ 5 km/s, typical of hyperbolic ejecta of the solar system. 

Adopting parameters from the minimum mass solar nebula (Weidenschilling 1977; Hayashi 1981), considering a range of planetesimal size distributions derived from observations of asteroids and KBOs and theoretical coagulation models (Bernstein et al.~2004; Bottke et al.~2005; Kenyon et al.~2008), and taking into account the results from Oort Cloud formation models (Brasser et al.~2011) for the fraction of planetesimals that are subject to weak escape from the early solar system, we estimated the number of meteoroids that may have been delivered to the weak stability boundary of the solar system over the lifetime of the Sun's birth cluster. Using this number and the probability of the weak capture of meteoroids on weakly escaping orbits, we calculated the number of weak transfer events from the early solar system to the nearest star in the cluster,  assuming it is a solar-type and harbors a planetary system. This number depends strongly on the adopted planetesimal size distribution.  We find that, for the cases where the power law size distribution at the small sizes ($dN/dD \propto D^{-q_2}$ for $D<D_0$) has index $q_2=3.5$, the expected number of weak transfer events between two solar type stars is of the order of  $10^{14}$--$ 3 \cdot 10^{16}$; for a shallow size distribution ($q_2=1.1$) the number is of the order of $10^{4}$.  

We conclude that solid material could have been transferred in significant quantity from the solar system to other solar-type stars in its birth cluster via the weak transfer mechanism described here. 

\subsection{Implications for lithopanspermia}
\label{sec:pans}

Section \ref{sec:summary} indicates that the transfer of planetesimals via the weak transfer mechanism described in this paper is likely to be the dominant process for the exchange of solid material amongst planetary systems in a star cluster. This is of interest for lithopanspermia because if life arose in any of these  systems before the cluster dispersed, this mechanism may have allowed the exchange of life-bearing rocks amongst the planetary systems in the cluster.  Within the context of the solar system's formation and dynamical history, in this section we discuss how much material originating from the Earth's crust may have been available for weak transfer (Section \ref{sec:biorelevant}), and, given the time constraints imposed by the weak transfer mechanism, whether or not there was a ``window of opportunity'' for lithopanspermia to take place (Section \ref{sec:time}). 

\subsubsection{Earth crustal material available for weak transfer}
\label{sec:biorelevant}
We first obtain an order-of-magnitude estimate of how much material may have been ejected from the Earth's crust as a consequence of the heavy bombardment that took place before the cluster dispersed. Following Adams and Spergel (2005), we assume that $l$ km of the Earth surface was ejected, with a total mass of $M_{\rm B} \sim 3\Big({l{\rm (km)} \over R_\oplus}\Big) M_\oplus \sim 5\cdot10^{-4} \cdot l{\rm (km)} M_\oplus \sim 3\cdot10^{24} \cdot l{\rm (km)}$ g. Adopting a power-law distribution, $dN/dm \propto m^{-\alpha}$, with $\alpha$ = 5/3, $m_1$ = 10 kg and $m_2=10^{-9}$ M$_\oplus$ (corresponding to objects 10 km in size), this total mass would be distributed in $N_B 
\sim {2-\alpha \over \alpha-1} {M_{\rm B} \over m_1^{\alpha-1}m_2^{2-\alpha}}
\sim 2\cdot10^{15} \cdot l{\rm (km)}$ bodies. A significant fraction of these fragments would have been ejected on hyperbolic orbits, i.e. beyond the domain of weak escape. To estimate how many of these bodies may have populated the weak stability boundary, we use the Oort Cloud formation efficiency of $\sim$ 1\% (Brasser et al. 2011): of the order of  $2\cdot10^{13} \cdot l{\rm (km)}$ bodies with a terrestrial origin that may have been subject to weak escape. 

An additional factor to consider is that a significant fraction of the Earth ejecta resulting from the bombardment would have been heated by shocks to pressures and temperatures high enough to sterilize the fragments (up to 50 GPa and several 100 $\degr$C respectively). However, a few percent of the material that originated from the spall region of the impacts --located a few projectile diameters away from the impact point in an area where the shocks cancel out-- would have remained weakly shocked, achieving a peak temperature $<$ 100 $\degr$C that would allow microorganisms to survive (Artemieva \& Ivanov 2004; Fritz et al. 2005, Pierazzo \& Chyba 1999, 2006).  In fact, laboratory experiments have confirmed that several microorganisms embedded in martian-like rocks have survived under shock pressures similar to those suffered by martian meteorites upon impact ejection (St\"offler et al. 2007; Horneck et al. 2008). Other laboratory experiments confirm that bacteria and yeast spores and microorganisms in a liquid can survive impacts with shock pressures of the order of GPas (Burchell et al. 2004; Willis et al. 2006; Hazell et al., 2010; Meyer et al. 2011).  Assuming only 1\% of the ejected Earth material remained weakly shocked, and factoring this into our estimate above, we get that $\sim 2\cdot10^{11} \cdot l{\rm (km)}$ life-bearing rocks with an Earth origin may have been subject to weak escape. 

Using the weak capture probability, $1.5\cdot10^{-3}$, derived in Section \ref{sec:mc_results}, we estimate that the total number of lithopanspermia events between the Earth and the nearest solar-type star in the cluster may have been of the order of $3 \cdot 10^{8} \cdot l{\rm (km)}$, where $l$ is the depth of the Earth's crust in km that was ejected during the ``window of opportunity".

\subsubsection{Time constraints}
\label{sec:time}

Now it is necessary to discuss the time constrains. We focus on two key aspects: (a) whether there is evidence that life may have arisen on Earth before the cluster dispersed, and (b) the survival of life to the hazards of outer space during the timescales relevant to weak transfer. 

\subsubsubsection{Is it possible that life arose on Earth before the cluster dispersed?}

The age of the solar system can be established from the dating of its oldest solids, the CAI inclusions in C-chondrites, that formed
$4.570\pm0.002$ Ga when the refractory elements in the solar nebula first started to condense at temperatures of approximately 2000 K (Lugmair \& Shukolyukov 2001). Let us assume that the stellar cluster was also born at that time. From Hf-W chronometry, it is estimated that the crystallization of the lunar magma oceans took place 4.527$\pm$0.010 Ga (Kleine et al. 2005), setting the time of the giant collision of a Mars-sized proto-planet with Earth that stripped part of its mantle and formed the Moon (Canup 2004). The detrital zircons found in Jack Hills in Western Australia show evidence that the Earth may have cooled down from this Moon-forming collision when the solar system was $\sim$ 70 Myr old; this evidence comes from the heterogeneity of the zircons Hf isotope ratio, $^{176}$Hf/$^{177}$Hf, a tracer of the crust/mantle differentiation (Harrison et al. 2005). Furthermore, the high oxygen isotope ratio, $^{18}$O/$^{16}$O, of 3.91--4.28 Gyr old Jack Hills detrital zircons suggests that the original rocks formed from magma containing recycled continental crust that had interacted with water near the surface. This indicates that liquid water was circulating in the upper crust of the Earth when the solar system was only 288 Myr old (Mojzsis et al 2001). Another study showed high oxygen isotope ratios in 4.404 Gyr old zircons, suggesting liquid water may have been present at an even earlier time, when the solar system was 164 Myr old (Wilde et al. 2001). The temperate conditions and possible presence of a hydrosphere indicate habitable conditions that may have allowed life to emerge during this period.

The carbon isotopic ratio of tiny inclusions of graphite in 3.85 Gyr old sedimentary rocks in Greenland show an increased $^{12}$CO/$^{13}$CO ratio that is indicative of biological activity, implying that life may have emerged before the solar system was 718 Myr old (Mojzsis et al 1996).  If this age estimate (based on uranium-lead dating of zircons) is correct, this means that life may have been extant very shortly after the end of the Late Heavy Bombardment (LHB).  Rather than abiogenesis taking place during such a short period, this favors the hypothesis that life emerged during the Hadean time and survived the LHB. In fact, in a study of the degree of thermal metamorphism suffered by the Earth's crust during the LHB, Abramov \& Mojzsis (2009) concluded that that it is unlikely that the entire crust was fully sterilized and that a microbial biosphere, if it existed, likely survived the LHB.  

In this section on the implications of weak transfer on the possibility of lithopanspermia, we will work with the hypothesis that life emerged during the Hadean time. This is motivated by the possible presence of oceans under temperate conditions and by the timescales for abiogenesis. Note, however, that evidence for life as early as 3.85 Ga as mentioned above is controversial (Moorbath 2005 and references therein). There is less controversial evidence of a sulfur-based bacterial ecosystem in Western Australian rocks with an age of approximately 3.4 Gyr, i.e. when the solar system was approximately 1170 Myr old (Wacey et al. 2011; see also Schopf et al. 1993). But this timeframe would be too late for lithopanspermia via weak transfer to take place because at this time the solar maternal cluster would have dispersed. The timescales for abiogenesis vary from 0.1--1 Myr for hydrothermal conditions at the deep sea, to 0.3--3 Myr for warm puddle conditions in shallow water, to 1--10 Myr for subaeric conditions in the soil, at least an order of magnitude less than the lifetime of the stellar cluster.  

If life arose on Earth shortly after there is evidence of liquid water on its crust, the ``window of opportunity'' for life-bearing rocks to be transferred to another planetary system in the cluster opens by the time liquid water was available, at 164--288 Myr, and ends by the cluster dispersal time, $T_{\rm cluster} \sim$ 135--535 Myr  (Adams 2010). Within this timeframe, there was a mechanism that allowed large quantities of rocks to be ejected from the Earth: the ejection of material resulting from the impacts at Earth during the heavy bombardment of the inner solar system. This bombardment period lasted from the end of the planet accretion phase until the end of the LHB 3.8 Ga, i.e. it finished when the solar system was approximately 770 Myr old (Tera et al.~1974; Mojzsis et al.~2001; Strom et al.~2005). It represents evidence that planetesimals were being cleared from the solar system several hundred million years after planet formation (Strom et al.~2005; Tsiganis et al.~2005; Chapman et al.~2007). This period of massive bombardment and planetesimal clearing encompassed completely the ``window of opportunity'' for the transfer of life-bearing rocks discussed above and therefore provides a viable ejection mechanism that may have led to weak transfer.

\subsubsubsection{Survival of life to the hazards of outer space during the timescales relevant to weak transfer}

A final consideration for lithopanspermia is the survival of microorganisms to the hazards of radiation during their long journey in outer space. Valtonen et al. (2009) used a computer model to account for the effects of Galactic cosmic rays from all elements up to nickel $Z$ = 28, and for the effects of natural radioactivity in meteorites characteristic of Earth and Mars. They found the following maximum total survival times in interstellar space (that depend on the size of the parent body):
12--15 Myr (for sizes of 0.00--0.03 m), 
15--40 Myr (0.03--0.67 m),
40--70 Myr (0.67--1.00 m),
70--200 Myr (1.00--1.67 m), 
200--300 Myr (1.67--2.00 m), 
300--400 Myr (2.00--2.33 m) and 
400--500 Myr (2.33--2.67 m). 
To put these lifetimes into context, note that under the best case scenario in which life emerged at the time liquid water was available in the upper crust (164 Myr or 270 Myr depending on the authors), lithopanspermia would have had a time window of up to 260--370 Myr (assuming the age of the cluster was in the upper end of $\sim$ 535 Myr). But the survival timescales above need to be compared, not to this time window, but to the transfer timescales associated with the weak transfer mechanism described in this paper. The latter are as follows: (a) {\it Timescale for ejection:} the numerical simulations carried by Melosh (2003) indicate that the minimum time between the ejection of meteorites from the Earth and exit from the solar system is 4 Myr, with a median time of 50 Myr.  Under our scenario, the time for a meteoroid to exit the solar system can be is estimated using Barker's equation, that gives the time of flight along a parabolic trajectory from periapsis with respect to the central star to the distance $R_{esc}(m_1)$. For a periapsis of 5 AU (at the location of the perturbing planet) and $R_{esc}$ =  1.8$\cdot$10$^5$ AU (corresponding to a solar-mass central star), we get that the exit timescale is $\sim$ 6 Myr. (2) {\it Timescale for interstellar transfer:} a meteoroid moving at the low velocity of 0.1 km/s (typical of the weak transfer mechanism) will take about 3--5 Myr to reach a neighboring star (located at a distance $D \approx 10^5$ AU). (3) {\it Timescale to land on a terrestrial planet:} Assuming that the neighboring star also harbors a planetary system, the weak capture mechanism would be active; a captured meteoroid would typically need to make multiple periapsis passages, i.e., some tens of millions of years, before collision with a planet.  We see therefore that the timescales for weak transfer compared to the microorganism survival timescales estimated by Valtonen et al. (2009) indicate that the survival of microorganisms could be viable via meteorites exceeding $\sim1$ m in size.

It is also of interest to study the exchange of prebiotic molecules between planetary systems, as they are more robust to the hazards of outer space. Simple amino acids like glycine have been found in several carbonaceous meteorites and in Stardust samples returned from comet Wild-2 (Elsila et al. 2009). Iglesias-Groth et al. (2011) argue that amino acids like glycine likely formed in the ISM and in chiral excess, and are therefore omnipresent\footnote{Iglesias-Groth et al.~(2011) study is based on the radiolysis and radioracemization rate constants derived from laboratory experiments in which glycine was exposed to doses similar to those delivered by the decay of natural radionuclides in comets and asteroids during 1 Gyr. The authors extrapolate to the solar system age and estimate the original concentration of amino acids at the time of solar system formation, concluding that "amino acids were formed in the interstellar medium and in chiral excess and then were incorporated in comets and asteroids at the epoch of the Solar System formation."}. This indicates that they may be available throughout the solar system, increasing the volume of material that may be subject to weak transfer. Moreover, there is no reason to assume that the fundamental hydrocarbon chemistry from which life developed was not present in other planetary systems at the time of their formation. Even though glycine has yet to be detected in the interstellar medium, Lattelais et al. (2011) points out that its non-detection (in spite of extensive radio surveys) is probably explained because neutral glycine is not the most stable isomer and therefore is not dominant. 

The discussion above assesses the possibility that life on Earth could have been transferred to other planetary systems when the Sun was still embedded in its stellar birth cluster. But could life on Earth have originated beyond the boundaries of our solar system? Our results indicate that, from the point of view of dynamical transport efficiency, life-bearing extra-solar planetesimals could have been delivered to the solar system via the weak transfer mechanism if life had a sufficiently early start in other planetary systems, before the solar maternal cluster dispersed. An early microbial biosphere, if it existed, likely survived the LHB. Thus, both possibilities remain open: that life was "seeded" on Earth by extra-solar planetesimals or that terrestrial life was transported to other star systems, via dynamical transport of meteorites.

\begin{center}
{\bf APPENDIX A: Earth-to-Moon weak transfer: study case for a large focussing factor}
\end{center}

As noted in Sections \ref{sec:oomprob} and \ref{sec:mc}, adopting for the focussing factor, $G_f = 1 + ({v_{\rm esc} \over v_{\infty}})^2$, a value of $G_f = 2$ is a conservative estimate. This factor can be substantially larger if at a given $R_{cap}(m_1^*)$, $v_{\infty}$ is  smaller than the approximate value of $U = 1$ km/s, while the value of $v_{esc}$ remains the same as $U$. Dynamically, the way to decrease $v_{\infty}$ with respect to $P_1^*$, as the meteoroid $P_0$ approaches $P_1^*$, is for $P_0$ to decrease its relative velocity. This process  has been shown to exist, for example, by the operational spacecraft {\it Hiten}, that was transferred from the Earth to the Moon on a trajectory that goes to ballistic capture at a given distance from the Moon.  Figure 5 shows a trajectory of the spacecraft leaving the Earth at a periapsis distance of $200$ km and going to  a periapsis distance of $500$ km from the Moon after 80 days, where the osculating eccentricity with respect to the Moon was 0.945. When it arrived at lunar periapsis, its velocity was approximately $v_{esc}$, while its $v_{\infty}$ went from a value of $1$ km/s to $0$ km/s (Belbruno 2004), implying $G_f = \infty$. Figure 6 shows the Kepler energy $E_k$ of the spacecraft with respect to the Moon along this transfer. At a sufficiently far distance from the Moon, where $v_{\infty} \approx \sqrt{2E_k}$, we see that $E_k \rightarrow 0$  implying $v_{\infty}\rightarrow 0$.


\begin{center}
{\bf APPENDIX B: Detailed results from the Monte-Carlo simulations}
\end{center}

For each case, the Monte Carlo simulation ran 5 million trajectories, a sufficient number so that the randomization of the initial values produces distributions of the parameters that span their respective ranges. The resulting distributions are shown in the histograms of Figures 7--21. Figures 7--11 correspond to Case 1, Figures 12--16 to Case 2 and Figures 17--21 to Case 3. The histograms are normalized by the total number of cases, so that they represent the probability density function of the parameter being measured (i.e. the integral under the curve is 1). Histograms labeled as ``all cases''  include all trajectories, regardless of whether or not capture is achieved, while histograms labeled as ``capture conditions" include only those cases that end in weak capture near $P_1^*$. We now describe some of the features of the histograms corresponding to Case 1. The other two cases are similar.  

Figure 7 shows that weak capture is inhibited for large initial separations between $P_0$ and $P_1$ (i.e. large $|{\bf{r_{0/1}}}|$). This is because if escape happens at a larger $|{\bf{r_{0/1}}}|$, it implies a larger v$_\infty$ and larger relative velocity with respect to $P_1^*$, decreasing the odds of capture. Note that the input distribution for $|{\bf{r_{0/1}}}|$ is non-uniform, as we are forcing the kinetic energy with respect to P$_1$ to be positive at $t = 0$ (which favors  higher values of $|{\bf{r_{0/1}}}|$, explaining the shape of the ``all cases" histogram); if we were to adopt  a uniform initial distribution for $|{\bf{r_{0/1}}}|$, we would get a decaying exponential or half-Gaussian distribution for the $|{\bf{r_{0/1}}}|$ of the capture cases.  

Figure 8 shows the initial velocity of  $P_0$ with respect to $P_1$, corresponding to the velocity distribution  in Figure 2. [Note that the latter figure shows $v_\infty$; to compare the two,  the escape velocity  ($\sim$ 0.2 km/s) needs to be subtracted from the abscissa of Figure 8].  The peak in the ``capture conditions" histogram around 1 km/s is due to the input distributions: the relative velocities of the two primaries were set to be less than 1 km/s, while the inertial velocities were forced to be less than 2 km/s.  Figure 9 shows that these various restrictions cause a peak in the distribution of initial inertial velocities of the primaries at $\sim$ 1 km/s. Therefore, the most likely velocity that would allow the particle to approach P$_1^*$ would  be $\sim$1 km/s.   The distribution of $v_\infty$  in Figure 2 extends out to 5 km/s; these larger velocities correspond to the small peak in the ``all cases" histogram at 2 km/s (in Figure 8); however, capture conditions become more unlikely for these higher velocities and therefore this peak is not present in the ``capture conditions" histogram. 


\begin{center} {\it Acknowledgments} \end{center}
E. B. acknowledges support from NASA Grant NNX09AK61G in the AISR Program of SMD. 
A. M.-M. acknowledges funding from the Spanish MICINN (Ram\'on y Cajal Program RYC-2007-00612, and grants AYA2009-07304 and Consolider Ingenio 2010CSD2009-00038). 
R.M. acknowledges support from NSF grant no.~AST-0806828.
Portions of this work by D.S. were performed under the auspices of the U.S.~Department of Energy by Lawrence Livermore National Laboratory under Contract DE-AC52-07NA27344. 
At Princeton University we would like to thank Robert Vanderbei for the use of his solar system simulation software, and David Spergel and Chris Chyba  for helpful discussion.

\clearpage 

\begin{center} {REFERENCES} \end{center}

\noindent Abramov, O. and Mojzsis, S. J.  (2009)
Microbial Habitability of the Hadean Earth During the Late Heavy Bombardment. 
Nature 459: 419--422. 

\noindent Adams, F. C., and Laughlin, G. (2001) 
Constraints on the Birth Aggregate of the solar system. 
Icarus 150: 151--162.

\noindent Artemieva, N., \& Ivanov, B. (2004)
Launch of martian meteorites in oblique impacts. 
Icarus 171: 84--101. 

\noindent Adams, F. C., and Spergel, D. N. (2005)
Lithopanspermia in Star Forming Clusters.
Astrobiology 5: 497--514.

\noindent Adams, F. C. (2010) 
The Birth Environment of the solar system.  
Annual Reviews of Astronomy and Astrophysics 48: 47--85. 

\noindent Belbruno, E. A. and Miller, J.  (1993) 
Sun-Perturbed Earth-to-Moon Transfers with Ballistic Capture. 
J.~Guid., Control and Dynamics 16: 770--775.

\noindent Belbruno, E. A. and Marsden, B. (1997)
Resonance Hopping in Comets. 
Astronomical Journal 113: 1433--1444.

\noindent Belbruno, E. A. (2004) 
{\em Capture Dynamics and Chaotic Motions in Celestial Mechanics}. 
Princeton University Press, Princeton. 

\noindent Belbruno, E. A. and Gott III, J. R. (2005) 
Where Did the Moon Come From?
Astronomical Journal 129: 1724--1745.

\noindent Belbruno, E. A. (2007) 
{\em Fly Me to the Moon: An Insider's Guide to the New Science of Space Travel}. 
Princeton University Press, Princeton. 

\noindent Belbruno, E. A., Gidea, M., and Topputo, F. (2010)  
Weak Stability Boundary and Invariant Manifolds. 
SIAM J. Applied Dynamical Systems 9: 1061--1089.

\noindent Bernstein, G. M., Trilling, D. E., Allen, R. L., Brown, M. E., Holman, M. and Malhotra, R. (2004) 
The Size Distribution of Trans-Neptunian Bodies. 
Astronomical Journal 128: 1364--1390.

\noindent Binney, J., and Tremaine, S. (1988) 
Galactic Dynamics, Princeton University Press, Princeton, US.

\noindent Booth, M., Wyatt, M. C., Morbidelli, A., Moro-Mart\'{i}n, A. and Levison, H. F. (2009)
The history of the solar system's debris disc: observable properties of the Kuiper belt.  
Monthly Notices of the Royal Astronomical Society 399: 385--398. 

\noindent Bottke, W. R., Durda, D. D., Nesvorny, D., Jedicke, R., Morbidelli, A., Vokrouhlicky D. and Levison, H. F. (2005) 
The Fossilized Size Distribution of the Main Asteroid Belt. 
Icarus: 175, 111--140.

\noindent Brasser, R., Duncan, M.~J., Levison, H.~F., Schwamb, M.~E., \& Brown, M.~E.\  (2011) 
Reassessing the formation of the inner Oort cloud in an embedded star cluster. 
Icarus, in press (arXiv:1110.5114). 

\noindent Burchell, M.~J., Mann, J.~R., \& Bunch, A.~W. (2004)
Survival of bacteria and spores under extreme shock pressures. 
Monthly Notices of the Royal Astronomical Society 352: 1273-1278. 

\noindent Canup, R.~M. (2004)
Simulations of a late lunar-forming impact. 
Icarus 168: 433--456.  

\noindent Chapman, C.~R., Cohen, B.~A. and Grinspoon, D.~H. (2007)
What are the real constraints on the existence and magnitude of the late heavy bombardment?
Icarus 189: 233--245

\noindent Dones, L., Gladman, B., Melosh, H. J., Tonks, W. B., Levison, H. F. and Duncan, M. (1999) 
Dynamical Lifetimes and Final Fates of Small Bodies: Orbit Integration vs Opik Calculations. 
Icarus 142: 509--524. 

\noindent Dones, L., Weissman, P.~R., Levison, H.~F. and Duncan, M.~J. (2004)
Oort cloud formation and dynamics.
In {\em Comets II}, edited by M. C. Festou, H. U. Keller and H. A. Weaver, University of Arizona Press, Tucson, pp. 153--174. 

\noindent Dormand, J.R., El-Mikkawy, A. and Prince, P.J. (1987) 
High-Order Embedded Runge-Kutta-Nystrom Formulae. 
Journal of Numerical Analysis 7: 423--430.

\noindent Elsila, J.~E., Glavin, D.~P., \& Dworkin, J.~P. (2009)
Cometary glycine detected in samples returned by Stardust. 
Meteoritics and Planetary Science 44: 1323--1330. 

\noindent Fritz, J., Artemieva, N., \& Greshake, A. (2005) 
Ejection of Martian meteorites. 
Meteoritics and Planetary Science, 40, 1393--1411. 

\noindent Garcia, R. and Gomez, G. (2007) 
A Note on the Weak Stability Boundary. 
Celestial Mechanics and Dynamical Astronomy 97: 87--100.

\noindent Gladman, B. J. (1997) 
Destination: Earth. Martian Meteorite Transfer. 
Icarus 130: 228--246.

\noindent Guckenheimer, J. and Holmes, P. (1983) 
{\em Nonlinear Oscillations, Dynamical Systems, and Bifucations of Vector Fields}, 
Springer Verlag. 

\noindent Harrison, T.~M., Blichert-Toft, J., M{\"u}ller, W., et al. (2005)
Heterogeneous Hadean Hafnium: Evidence of Continental Crust at 4.4 to 4.5 Ga.  
Science 310: 1947--1950.  

\noindent Hayashi, C. (1981) 
Structure of the Solar Nebula, Growth and Decay of Magnetic Fields and Effects of Magnetic and Turbulent Viscosities on the Nebula. 
Progress of Theoretical Physics Supplement 70: 35--53.

\noindent Hazell PJ, Beveridge C, Groves K, Appleby-Thomas G. (2010)
The shock compression of microorganism-loaded broths and emulsions: Experiments and simulations. 
International Journal of Impact Engineering 37: 433Ð440. 

\noindent Hernquist, L. (1990) 
An Analytical Model for Spherical Galaxies and Bulges. 
Astrophysical Journal 356: 359--364.

\noindent Horneck, G. (1993) 
Responses of Bacillus Subtilis Spores to Space Environment: Results from Experiments in Space. 
Origins of Life and Evolution of the Biosphere 23: 37-52.  

\noindent Horneck, G., St\"offler, D., Ott, S., Hornemann, U., Cockell, C. S. et al. (2008)
Microbial Rock Inhabitants Survive Hypervelocity Impacts on Mars-Like Host Planets: First Phase of Lithopanspermia Experimentally Tested. 
Astrobiology 8: 17--44.

\noindent Iglesias-Groth, S., Cataldo, F., Ursini, O., \& Manchado, A. (2011) 
Amino acids in comets and meteorites: stability under gamma radiation and preservation of the enantiomeric excess. 
Monthly Notices of the Royal Astronomical Society 410: 1447--1453.

\noindent Keyon, S. J., Bomley, B. C., O'Brien, D. P. and Davis, D. R. (2008)
Formation and Collisional Evolution of Kuiper Belt Objects. 
In {\em The Solar System Beyond Neptune}, edited by A. Barucci, H. Boehnhardt, D.
Cruikshank and A. Morbidelli, University of Arizona Press, Tucson, pp. 293--313.

\noindent Kleine, T., Palme, H., Mezger, K., \& Halliday, A.~N. (2005)
Hf-W Chronometry of Lunar Metals and the Age and Early Differentiation of the Moon.
Science, 310, 1671--1674. 

\noindent Lada, C.~J., \& Lada, E.~A. (2003)
Embedded Clusters in Molecular Clouds. 
Annual Review of Astronomy \& Astrophysics 41: pp.57--115. 

\noindent Lattelais, M., Pauzat, F., Pilm{\'e}, J., Ellinger, Y., \& Ceccarelli, C. (2011)
About the detectability of glycine in the interstellar medium.
 Astronomy \& Astrophysics 532: A39 

\noindent Levison, H.~F., Duncan, M.~J., Brasser, R., \& Kaufmann, D.~E. (2010)
Capture of the Sun's Oort Cloud from Stars in Its Birth Cluster. 
Science 329: 187. 

\noindent Lugmair, G.~W., \& Shukolyukov, A. (2001) 
Early solar system events and timescales. 
Meteoritics and Planetary Science 36: 1017--1026. 

\noindent Marsden, J. and Ross, S. (2006) 
New Methods in Celestial Mechanics and Mission Design.
 Bull. Amer. Math. Soc. 43: 43.

\noindent McSween, H.~Y., Jr.\ (1976)
A new type of chondritic meteorite found in lunar soil. 
Earth and Planetary Science Letters 31: 193-199. 

\noindent Melosh, H. J.  (2003) 
Exchange of Meteorites (and Life?) Between Solar Systems.
Astrobiology 3: 207--215.

\noindent Meylan, G., \& Heggie, D.~C. (1997)
Internal dynamics of globular clusters. 
The Astronomy and Astrophysics Review, 8: 1--143. 

\noindent Meyer, C., Fritz, J., Misgaiski, M., et al.\ (2011) 
Shock experiments in support of the Lithopanspermia theory: The influence of host rock composition, temperature, and shock pressure on the survival rate of endolithic and epilithic microorganisms. 
Meteoritics and Planetary Science, 46, 701--718.  

\noindent Mojzsis, S.~J.,  Arrhenius, G., McKeegan, K.~D., et al. (1996)
Evidence for life on Earth before 3,800 million years ago. 
Nature 384: 55--59. 

\noindent Mojzsis, S.~J., Harrison, T.~M. and Pidgeon, R.~T. (2001)
Oxygen-isotope evidence from ancient zircons for liquid water at the Earth's surface 4,300Myr ago. 
Nature 409: 178--181. 

\noindent Moro-Mart\'{\i}n, A. and Malhotra, R. (2005)  
Dust Outflows and Inner Gaps Generated by Massive Planets in Debris Discs. 
Astrophysical Journal 633: 1150--1167.

\noindent Pierazzo, E., \& Chyba, C.~F. (1999) 
Amino acid survival in large cometary impacts. 
Meteoritics and Planetary Science 34: 909--918. 

\noindent Pierazzo, E., \& Chyba, C.~F. (2006)
Impact Delivery of Prebiotic Organic Matter to Planetary Surfaces. 
Comets and the Origin and Evolution of Life,  Advances in Astrobiology and Biogeophysics, Springer, 137.  

\noindent Racca, G. D. (2003)
New Challenges to Trajectory Design by the Use of Electric Propulsion and Other Means of Wandering in the solar system. 
Celestial Mechanics and Dynamical Astronomy 85: 1--24.

\noindent Schopf, J. W. (1993)
Microfossils of the Early Archean Apex Chert: New Evidence of the Antiquity of Life. 
Science 260: 640--646. 

\noindent Schr{\"o}der, C., Rodionov, D.~S., McCoy, T.~J., et al. (2008)
Meteorites on Mars observed with the Mars Exploration Rovers
Journal of Geophysical Research (Planets) 113: 6.  

\noindent Smale, S. (1967)
Differential Dynamical Systems. 
Bull. AMS 73: 747-817.

\noindent St\"offler, D., Horneck, G., Ott, S., Hornemann, U., Cockell, C. S. et al. (2007) 
Experimental Evidence for the Potential Impact Ejection of Viable Microorganisms from Mars and Mars-Like Planets. 
Icarus 186: 585--588.

\noindent Strom, R.~G., Malhotra, R., Ito, T., Yoshida, F., \& Kring, D.~A. (2005) 
The Origin of Planetary Impactors in the Inner Solar System. 
Science 309: 1847 1850. 

\noindent Tera, F., Papanastassiou, D.~A. and Wasserburg, G.~J. (1974)
Isotopic evidence for a terminal lunar cataclysm. 
Earth and Planetary Science Letters 22: 1. 

\noindent Tsiganis, K., Gomes, R., Morbidelli, A., \& Levison, H.~F. (2005)
Origin of the orbital architecture of the giant planets of the Solar System. 
Nature 435: 459--461. 

\noindent Valtonen, M., Nurmi,  P., Zheng, J.-Q., et al. (2009) 
Natural Transfer of Viable Microbes in Space from Planets in Extra-Solar Systems to a Planet in our Solar System and Vice Versa. 
Astrophysical Journal 690: 210--215.  

\noindent Wacey, D., Kilburn, M. R., Saunders, M., Cliff, J. and Brasier, M. D. (2011)
Microfossils of sulphur-metabolizing cells in 3.4-billion-year-old rocks of Western Australia. 
Nature Geosciences, DOI: 10.1038/NGEO1238. 

\noindent Weidenschilling, S. J. (1977) 
The Distribution of Mass in the Planetary System and Solar Nebula. 
Astrophys. Space Sci. 51: 153--158.

\noindent Wilde, S.~A., Valley, J.~W., Peck, W.~H., \& Graham, C.~M. (2001)
Evidence from detrital zircons for the existence of continental crust and oceans on the Earth 4.4Gyr ago. 
Nature 409: 175--178 

\noindent Willis, M.~J., Ahrens, T.~J., Bertani, L.~E., \& Nash, C.~Z. (2006) 
BugbusterÑsurvivability of living bacteria upon shock compression. 
Earth and Planetary Science Letters 247: 185--196. 

\noindent Xia, Z.  (1992) 
Melnikov Method and Transversal homoclinic Points in the Restricted Three-Body problem.
JDE 96: 170.

\clearpage

\begin{deluxetable}{ccccc}
\tablewidth{0pc}
\tablecaption{Weak capture probability from Monte Carlo simulations$^{a}$}
\tablehead{
\colhead{Case} & 
\colhead{Mass of source} & 
\colhead{Mass of target} & 
\colhead{Number of} & 
\colhead{Weak Capture}\\
\colhead{} & 
\colhead{star $m_1$  (M$_{\odot}$)} & 
\colhead{star $m_1^*$  (M$_{\odot}$)} & 
\colhead{trajectories} & 
\colhead{Probability}}
\startdata
1 & 1    & 1      &  5$\cdot10^6$ & 0.15\%\\
2 & 1    & 0.5   &  5$\cdot10^6$ & 0.05\%\\ 
3 & 0.5 & 1      &  5$\cdot10^6$ & 0.12\%\\
\enddata
\tablenotetext{a}{All cases assume a cluster with N = 4300 members and cluster properties described in Section \ref{sec:clusprop}. }
\end{deluxetable}

\clearpage

\begin{deluxetable}{lcccccc}
\tablewidth{0pc}
\tablecaption{Estimated number of weak transfer events between the Sun and its closest cluster neighbor\tablenotemark{a}}
\tablehead{
\colhead{Case} & 
\colhead{$q_1$} & 
\colhead{$q_2$} & 
\colhead{$D_0$} & 
\colhead{N$_{D>26cm}$\tablenotemark{b}} &
\colhead{N$_{\rm R}$\tablenotemark{c}} & 
\colhead{N$_{\rm WTE} (m_1^* = 1)$\tablenotemark{d}}\\
\colhead{} & 
\colhead{} & 
\colhead{} &
\colhead{(km)} & 
\colhead{} & 
\colhead{} & 
\colhead{(N=4300)}
}

\startdata
A & 4.3 & 3.5 & 100		& 1.8$\cdot10^{21}$		& 1.8$\cdot10^{19}$  & 2.7$\cdot10^{16}$	\\
B & 3.3 & 3.5 & 2		      & 2.7$\cdot10^{20}$		& 2.7$\cdot10^{18}$  & 4.0$\cdot10^{15}$\\
C & 2.7 & 3.5 & 2 		& 8.1$\cdot10^{18}$		& 8.1$\cdot10^{16}$  & 1.2$\cdot10^{14}$\\
D & 4.3 & 1.1 & 100		& 5.5$\cdot10^{8}$		& 5.5$\cdot10^{6}$    & 8.2$\cdot10^{3}$\\
\enddata
\tablenotetext{a}{Considering four different cases (A, B, C, D) for the planetesimal size distribution. The size distribution is a broken power-law given by $dN/dD \propto$ D$^{-q_1}$ for $D>D_0$ and dN/dD $\propto$ D$^{-q_2}$ if D$<$D$_0$, with 
D$_{\rm max}$ = 2000 km ($\sim$ Pluto's size) and D$_{\rm min}$ = 1 $\mu$m ($\sim$ dust blow-out size).}
\tablenotetext{b}{N$_{D>26cm}$ is the total number of planetesimals with a diameter $D > 26$ cm , equivalent to masses $>10$ kg (assuming $\rho$ = 1 g/cm$^{-3}$).}
\tablenotetext{c}{N$_{\rm R}$ is the expected number of planetesimals with a diameter $D > 26$ cm  that populated the primordial Oort Cloud, estimated to be $\sim$1\% of  N$_{D>26cm}$.}
\tablenotetext{d}{N$_{\rm WTE} (m_1^* = 1)$ is the number of weak transfer events between the Sun and a neighboring solar-type cluster star with $m_1^*$ = 1 M$_{\odot}$ and assuming the cluster has N = 4300 members; this is calculated multiplying N$_{\rm R}$ by the weak capture probability in Table 1 corresponding to $m_1^*$ = 1 M$_{\odot}$, i.e.  N$_{\rm R}$ $\times$1.5$\cdot 10^{-3}$.}
\end{deluxetable}

\clearpage

\begin{center} {\it FIGURE CAPTIONS}\end{center}

\begin{center}
\includegraphics[scale=0.50,angle=0]{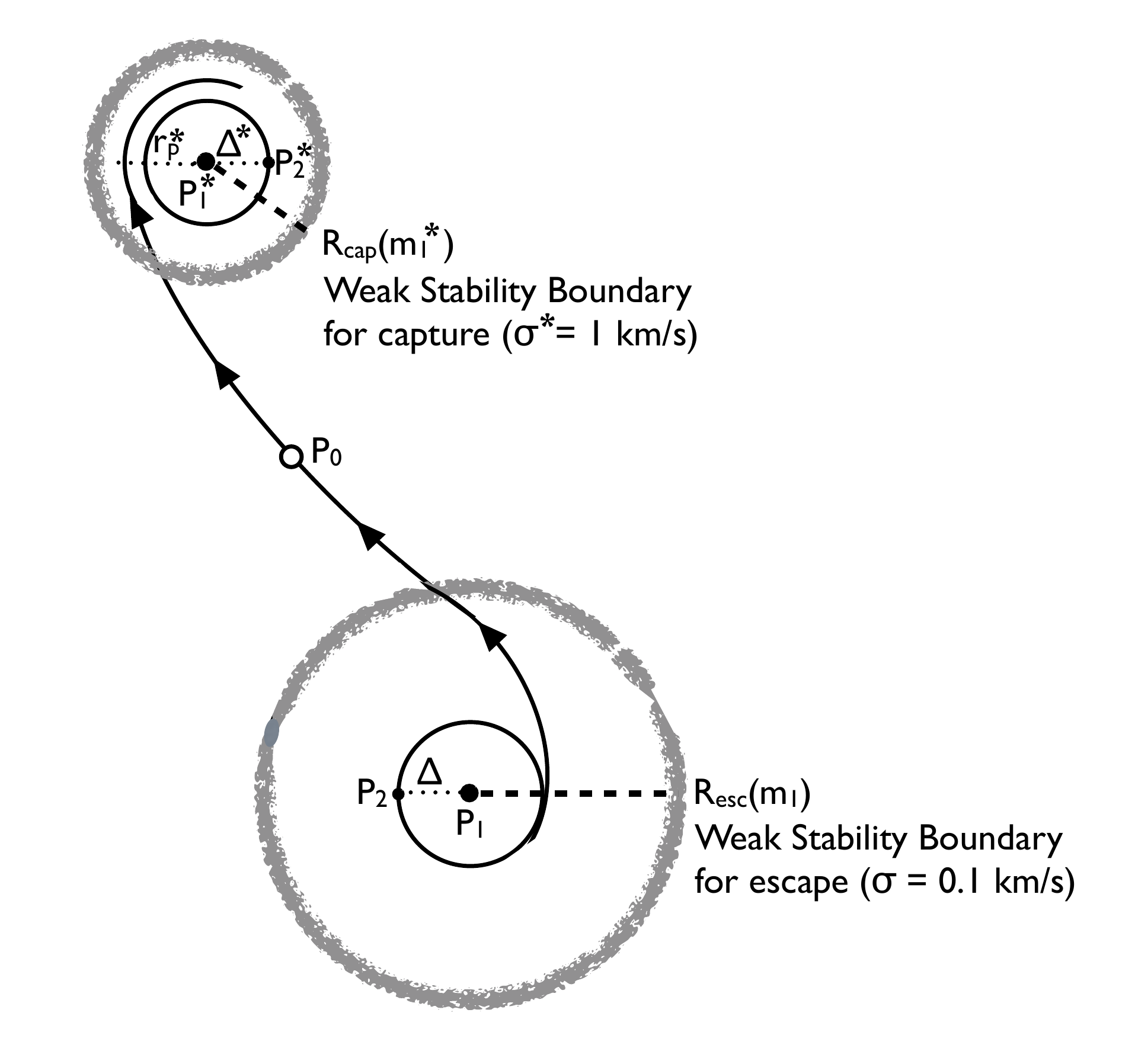}
\end{center}

{\bf Fig. 1.} Schematic representation of the weak transfer process. It consists on a meteoroid weakly escaping from a planetary system, and its subsequent weak capture by a neighboring planetary system in the stellar cluster. The meteoroid $P_0$ flybys planet $P_2$ and weakly escapes the central star $P_1$ at a distance $R_{\rm esc}(m_1)$.  $R_{\rm esc}(m_1)$ is approximately where the the weak stability boundary of $P_1$ is located, caused by the gravitational perturbation of the other $N-1$ stars in the cluster. The motion in this region is chaotic and lies in the transition between capture and escape from $P_1$. The meteoroid $P_0$ is then weakly captured by the neighboring cluster star  $P_1^*$ at a distance $R_{\rm cap}(m_1^*)$, moving to periapsis $r_p^*$ with respect to $P_1^*$ (motion projected onto a plane). The numerical computation pieces together trajectories from flyby of $P_2$ to the distance  $R_{\rm esc}(m_1)$ from $P_1$ using $RP3D$  and from $R_{\rm esc}(m_1)$ to the distance $R_{\rm cap}(m_1^*)$ from $P_1^*$. It has been demonstrated that the piecing together of solutions of two different three-body problems at $WSB(P_1)$ and $WSB(P_1^*)$ can be done in a well defined  and smooth manner (see Belbruno, 2004;  Marsden \& Ross, 2006).
\label{figure:transfernew}
 
\begin{center}
\includegraphics[scale=0.50,angle=0]{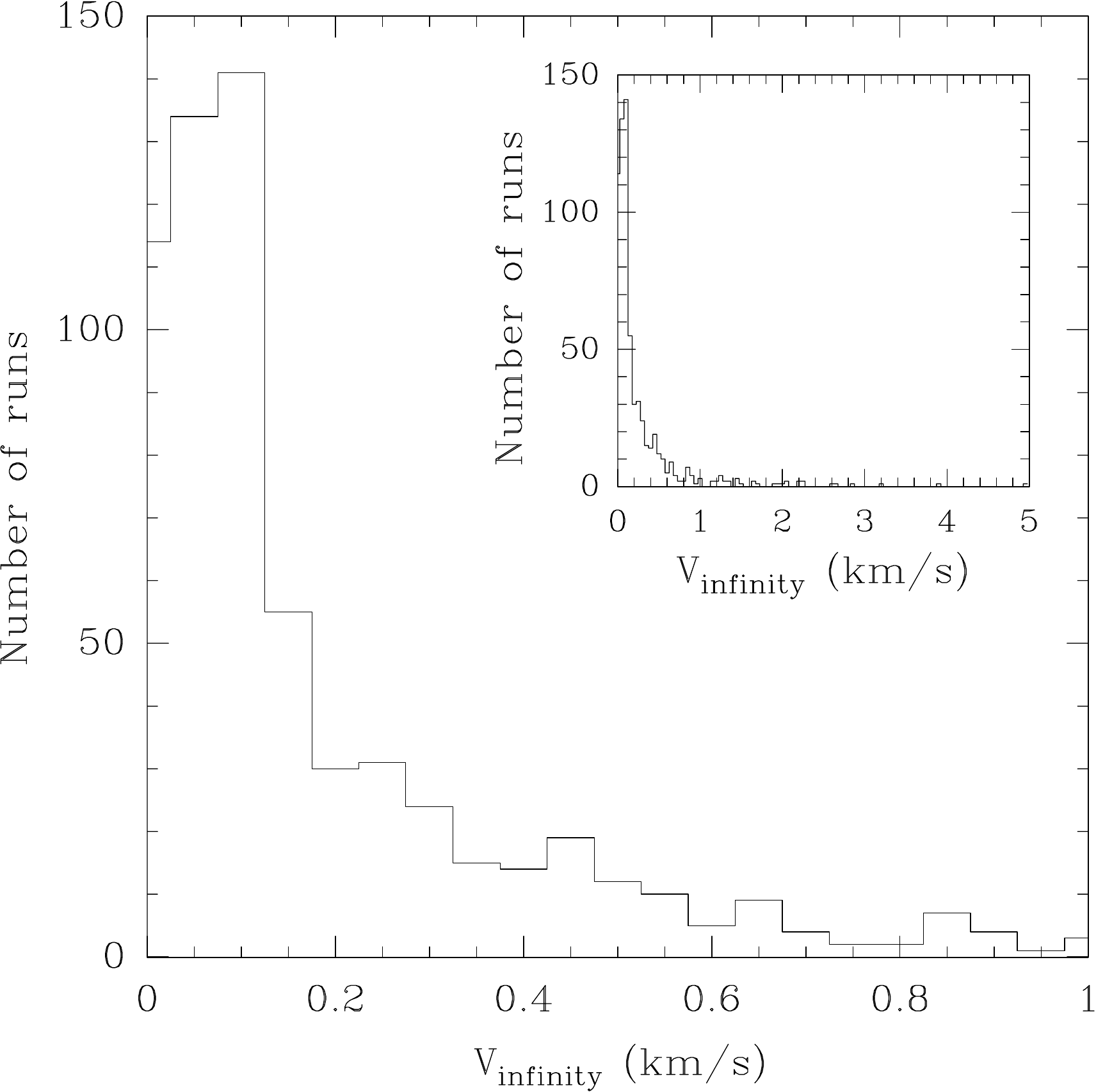}
\end{center}

{\bf Fig. 2.} Velocity distribution of 670 low-velocity test particles escaping from the solar system. The numerical model is the planar circular restricted three-body problem of the Sun, Jupiter and a massless particle.
\label{figure:vhist}

\begin{center}
\includegraphics[scale=0.50,angle=0]{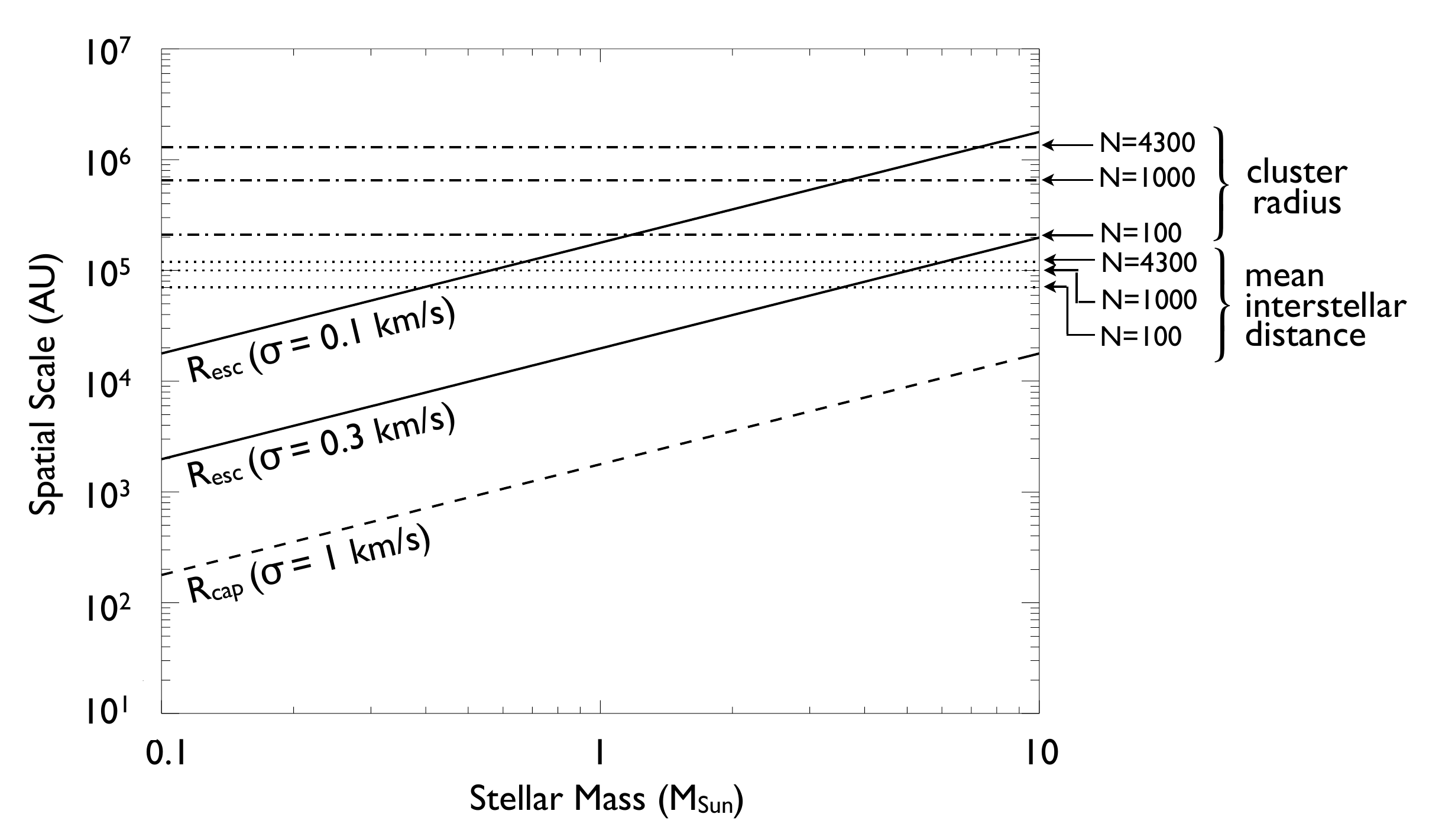}
\end{center}

{\bf Fig. 3.} The diagonal solid lines plot the spatial scale of the weak stability boundary for escape, $R_{esc}(m_1)=2Gm_1/\sigma^2$, as a function of stellar mass, $m_1$, for two values of the escape velocity, $\sigma=0.1$ and 0.3 km/s. The diagonal dashed line plots the weak stability boundary for capture, $R_{cap}(m_1^*)=2Gm_1^*/U^2$, for the average relative velocity of stars in the cluster, $U=1$ km/s. The horizontal dashed-dotted lines indicate the spatial scale of star clusters, $R_{cluster}$, while the horizontal dotted lines indicate the mean interstellar distance, D, for clusters consisting of N=100, N=1000 and N = 4300 members.
\label{figure:escape}

\begin{center}
\includegraphics[scale=0.50,angle=0]{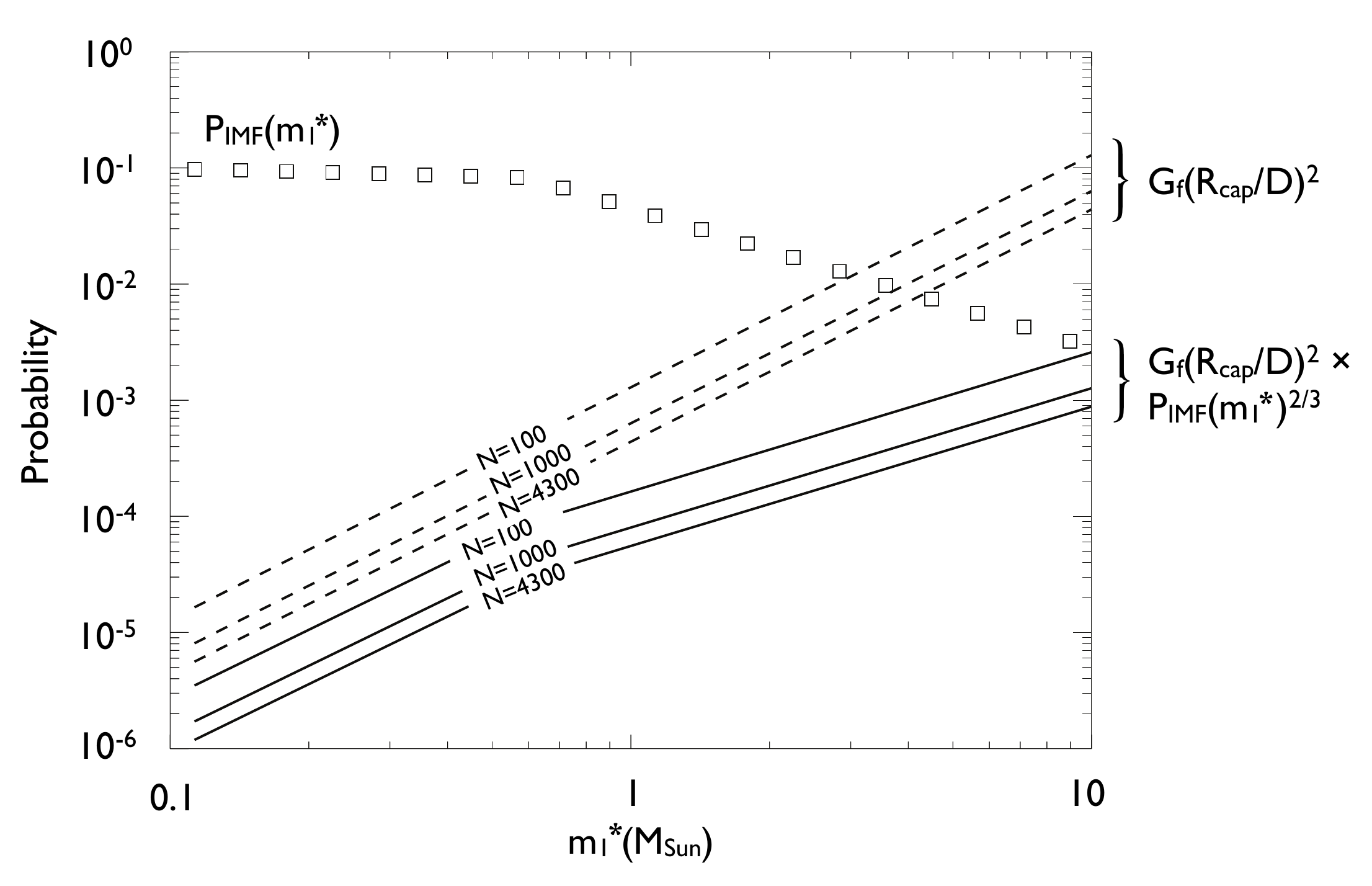}
\end{center}

{\bf Fig. 4.} The diagonal dashed lines plot the probability for weak capture of meteoroids, $G_f(R_{cap}/D)^2$, as a function of the target stellar mass ($m_1^*$), for three values of the number of stars in a star cluster. For $G_f$, we take the conservative value of 2. The solid lines plot the probability for weak capture of meteoroids by a neighbor star of stellar mass equal to that of the source star ($m_1=m_1^*$). The symbols indicate the stellar mass distribution.
\label{figure:imfcor}

\begin{center}
\includegraphics[scale=0.50,angle=0]{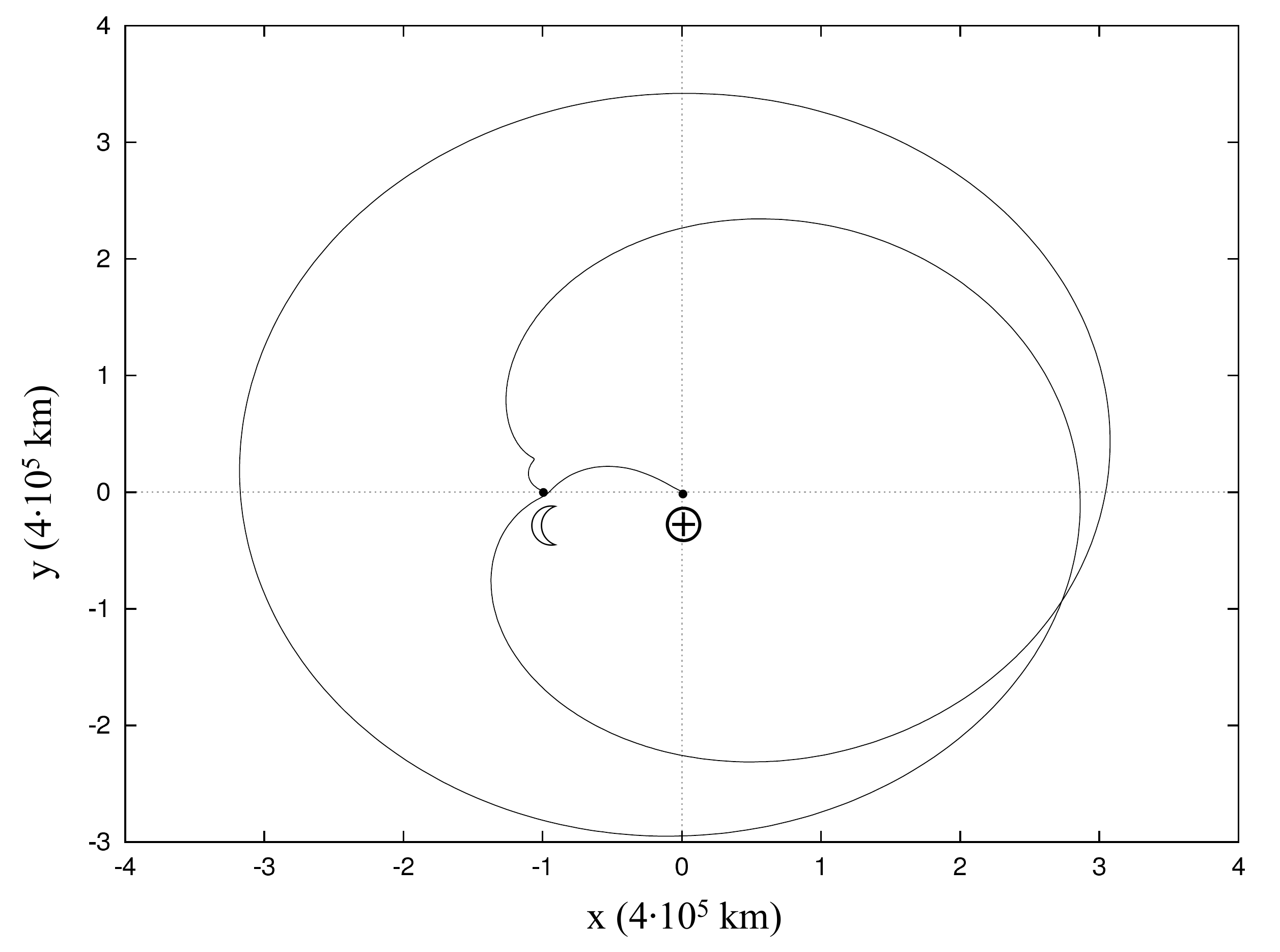}
\end{center}

{\bf Fig. 5.} Weak transfer of the spacecraft {\it Hiten} from the Earth (located at (0,0)) to the Moon (located at (-1,0)) via the lunar weak stability boundary. This is an Earth-Moon fixed rotating coordinate system, projected onto the Earth-Moon plane. The time of travel is 80 days, 14 hours. The x-axis and y-axis are in units of 4$\times$10$^{5}$~km.
\label{Figure:WSBTransfer}

\begin{center}
\includegraphics[scale=0.50,angle=0]{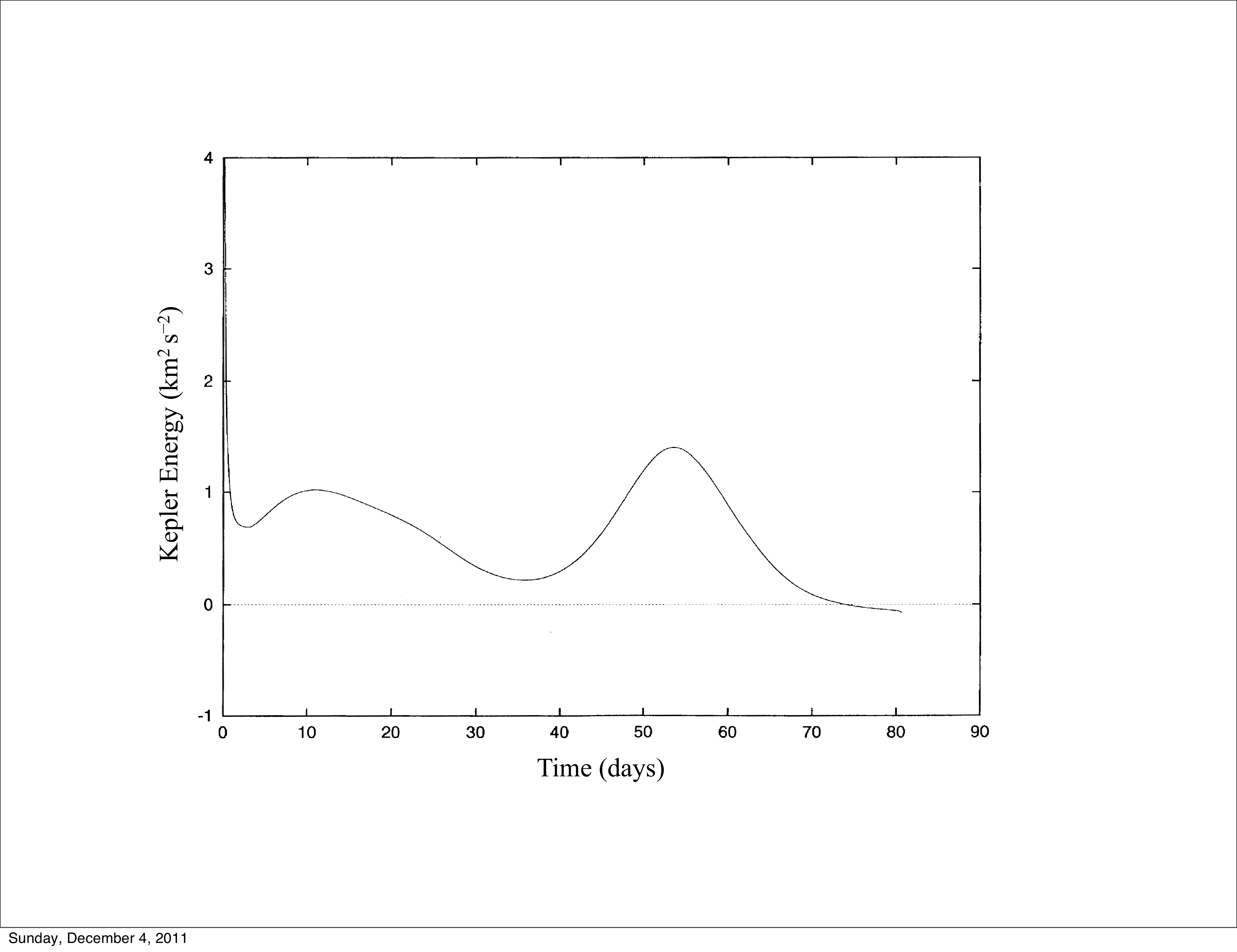}
\end{center}

{\bf Fig. 6.} Kepler energy of the spacecraft Hiten with respect to the Moon (in units of km$^2$s$^{-2}$) as a function of time (in units of days) along the trajectory shown in Fig. 5. 
\label{Figure:VinfPlot}

\begin{center}
\includegraphics[scale=0.50,angle=0]{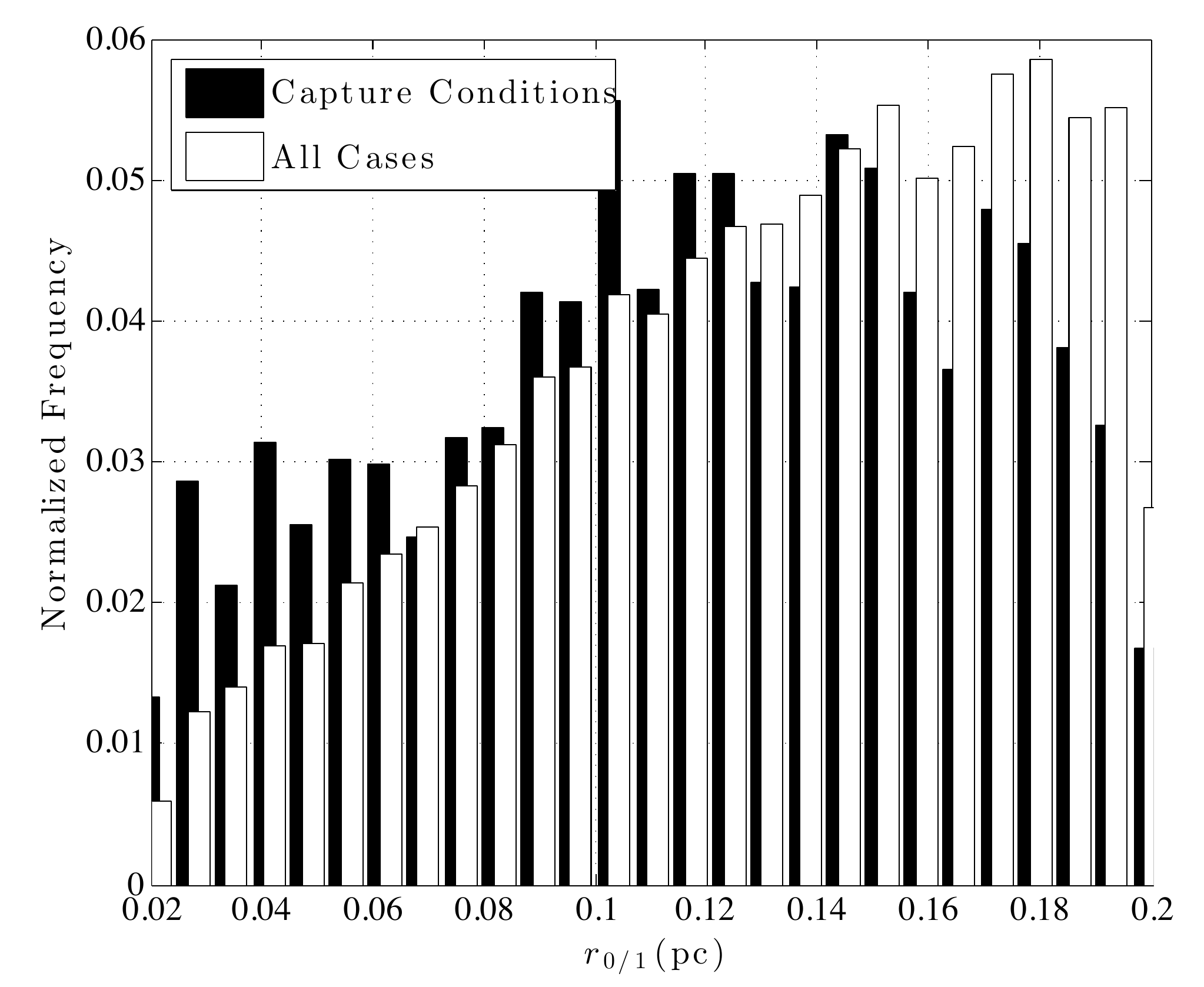}
\end{center}

{\bf Fig. 7.} Results from Monte Carlo simulations of 5 million trajectories between a star $P_1$ of mass $m_1 = 1$ M$_\odot$ and star $P_1^*$ of mass $m_1^* = 1$ M$_\odot$ (Case 1). The Figure shows the probability density function (normalized to 1) of the initial separation between $P_0$ and $P_1$ ($r_{0/1}$ = $|{\bf{r_{0/1}}}|$ at $t=0$).  
\label{figure:c1sepP0P1}

\begin{center}
\includegraphics[scale=0.50,angle=0]{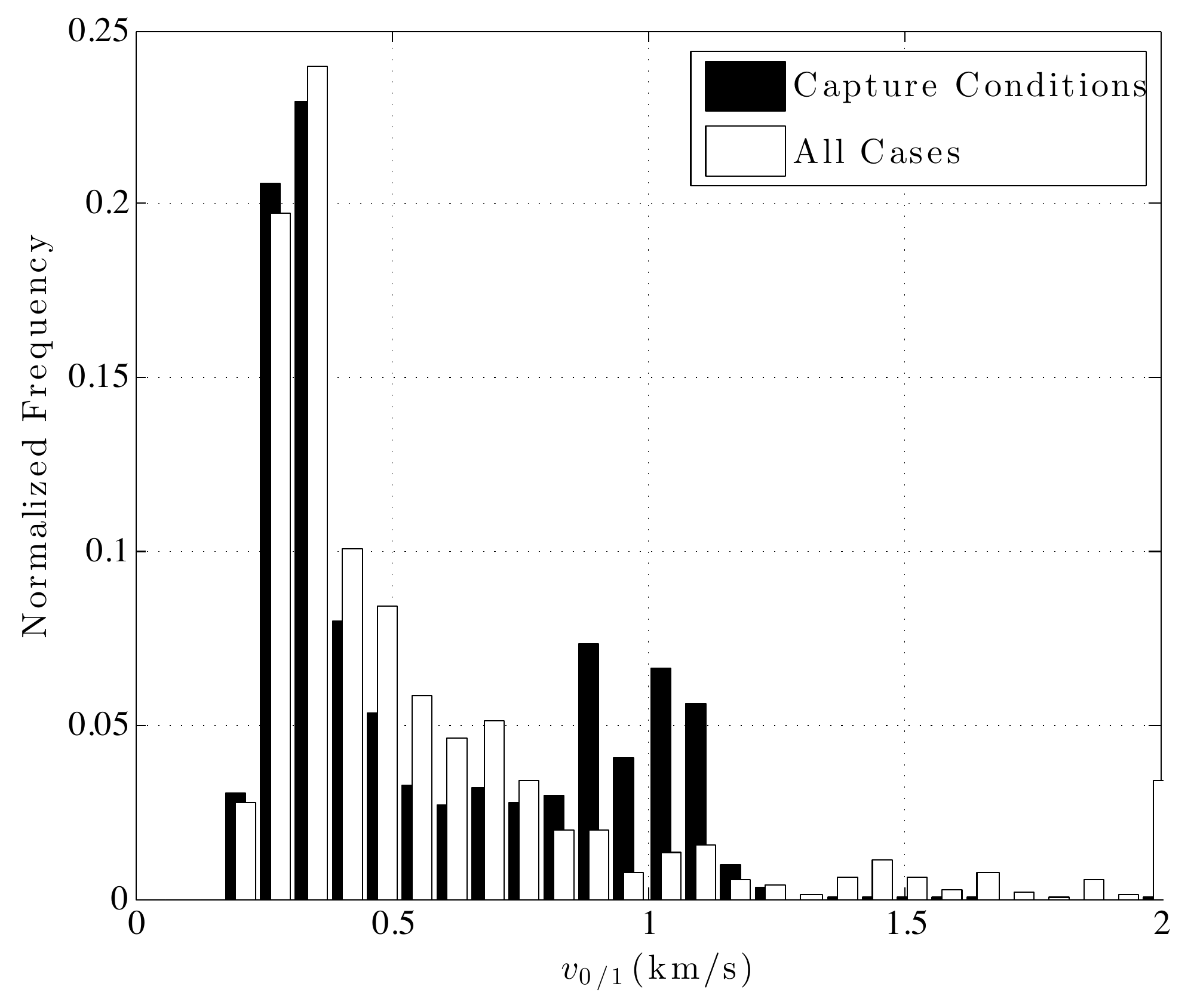}
\end{center}

{\bf Fig. 8.} Results from Monte Carlo simulations of 5 million trajectories between a star $P_1$ of mass $m_1 = 1$ M$_\odot$ and star $P_1^*$ of mass $m_1^* = 1$ M$_\odot$ (Case 1). The Figure shows the probability density function (normalized to 1) of the initial velocity of  $P_0$ with respect to $P_1$ ($v_{0/1}$ = $|{\bf{v_{0/1}}}|$ at $t=0$). 
\label{figure:c1velP0P1}

\begin{center}
\includegraphics[scale=0.50,angle=0]{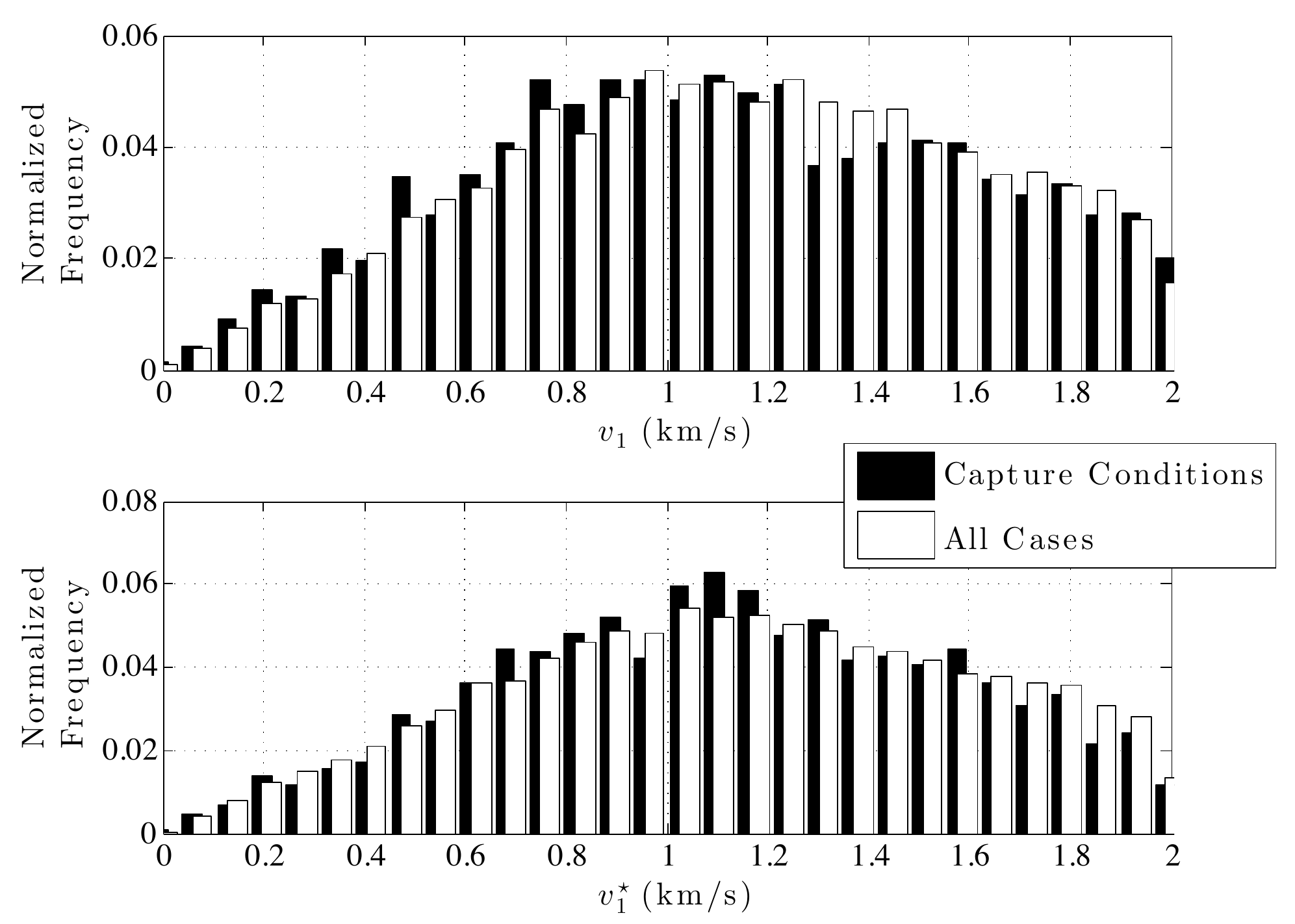}
\end{center}

{\bf Fig. 9.} Results from Monte Carlo simulations of 5 million trajectories between a star $P_1$ of mass $m_1 = 1$ M$_\odot$ and star $P_1^*$ of mass $m_1^* = 1$ M$_\odot$ (Case 1). The Figure shows the probability density function (normalized to 1) of the initial inertial velocities of $P_1$ and $P_1^*$ with respect to the cluster center ($v_1 = |{\bf{v_1}}|$ and $v_1^* = |{\bf{v_1^*}}|$ at $t=0$). 
\label{figure:c1velP1P1}

\begin{center}
\includegraphics[scale=0.50,angle=0]{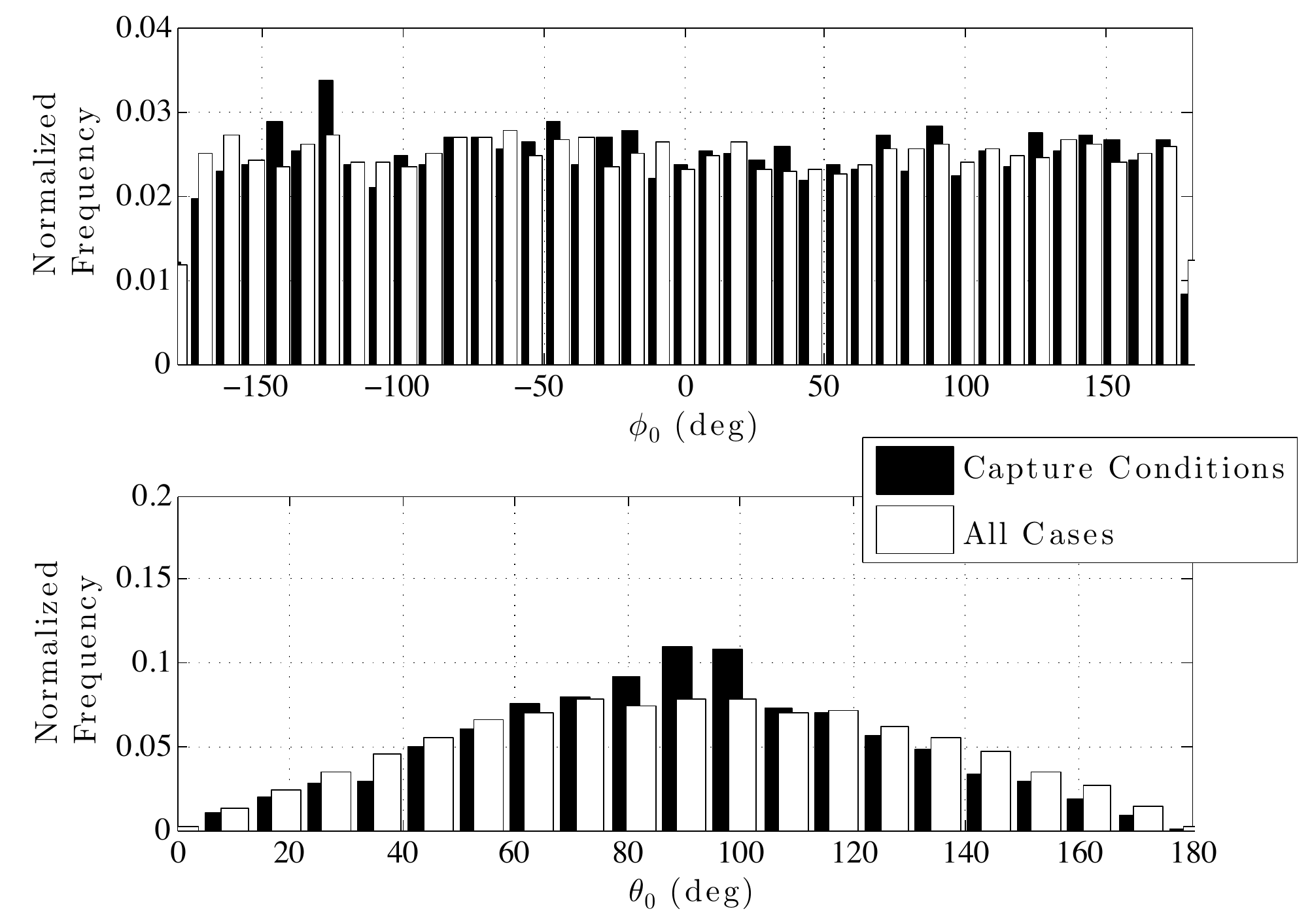}
\end{center}

{\bf Fig. 10.} Case 1: Probability density function of the initial orientation of $P_0$ about $P_1$ ($\phi_0$ and $\theta_0$ are the spherical angles). The integral under the curve is 1. 
\label{figure:c1oriP0P1}

\begin{center}
\includegraphics[scale=0.50,angle=0]{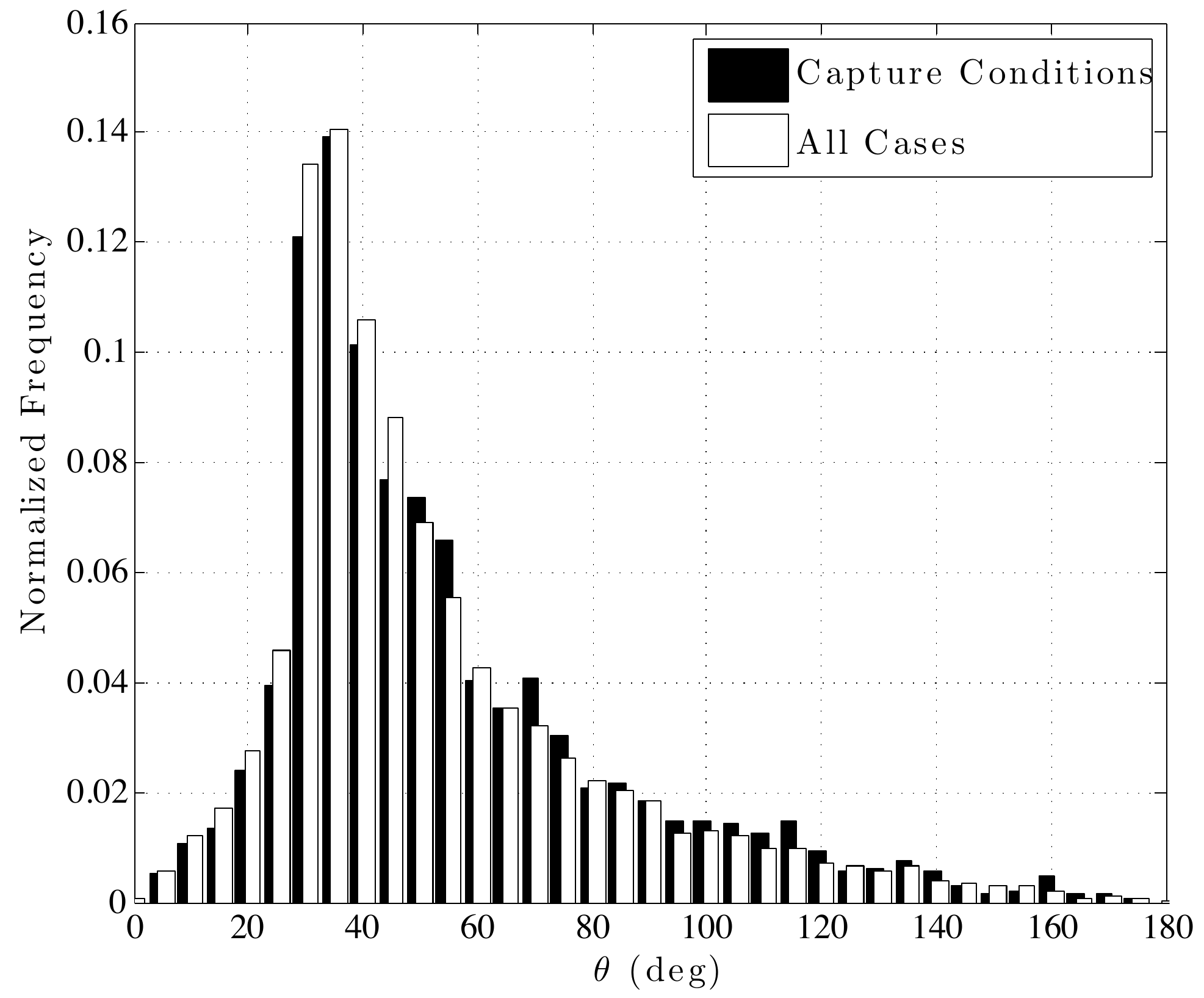}
\end{center}

{\bf Fig. 11.} Case 1: Probability density function of the angular separation between initial velocities of $P_1$, $P_1^*$. The integral under the curve is 1. 
\label{figure:c1oriP1P1}

\begin{center}
\includegraphics[scale=0.50,angle=0]{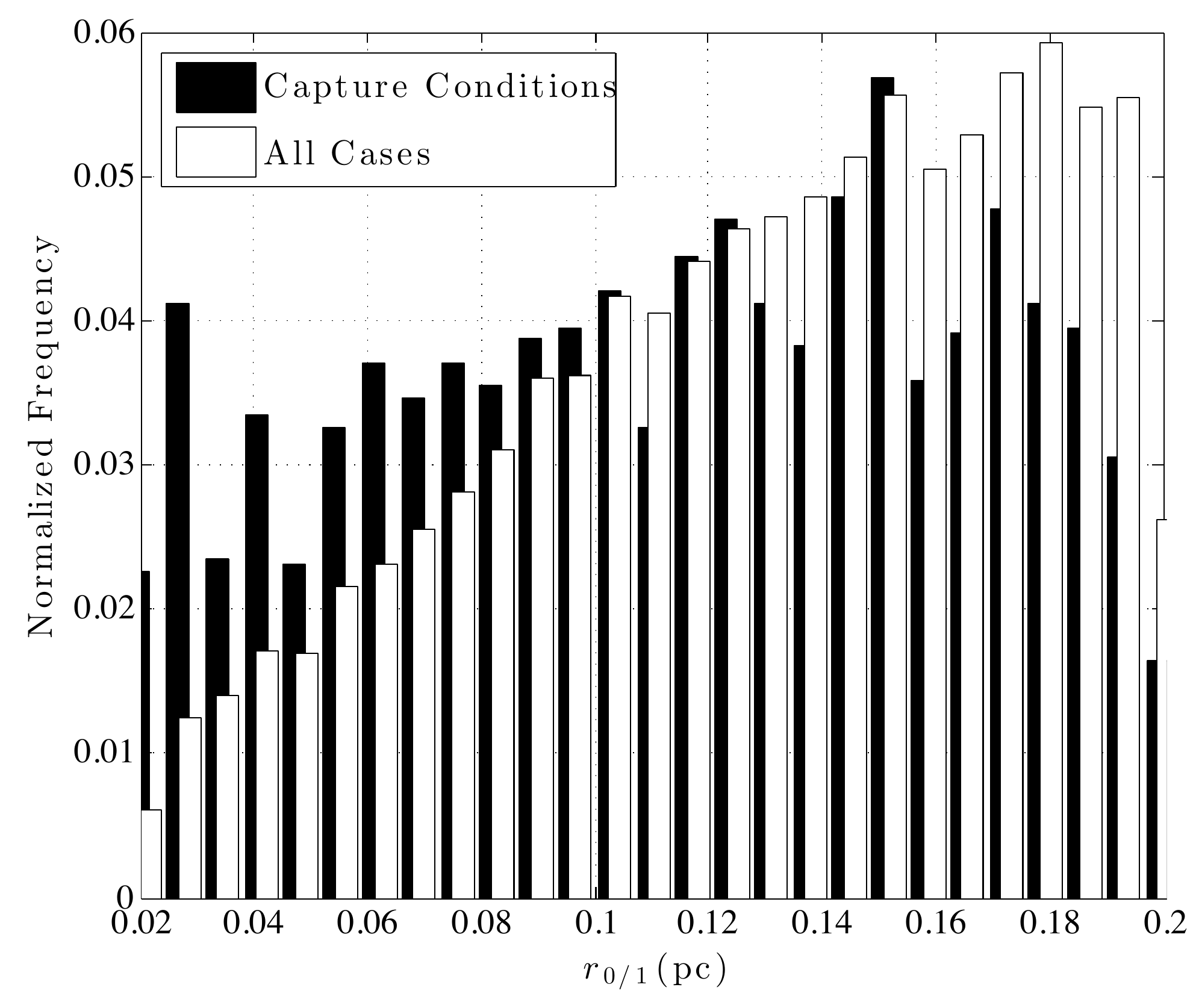}
\end{center}

{\bf Fig. 12.} Same as Figure 7 but for Case 2, with $m_1 = 1$ M$_\odot$ and $m_1^* = 0.5$ M$_\odot$.  
\label{figure:c2sepP0P1}

\begin{center}
\includegraphics[scale=0.50,angle=0]{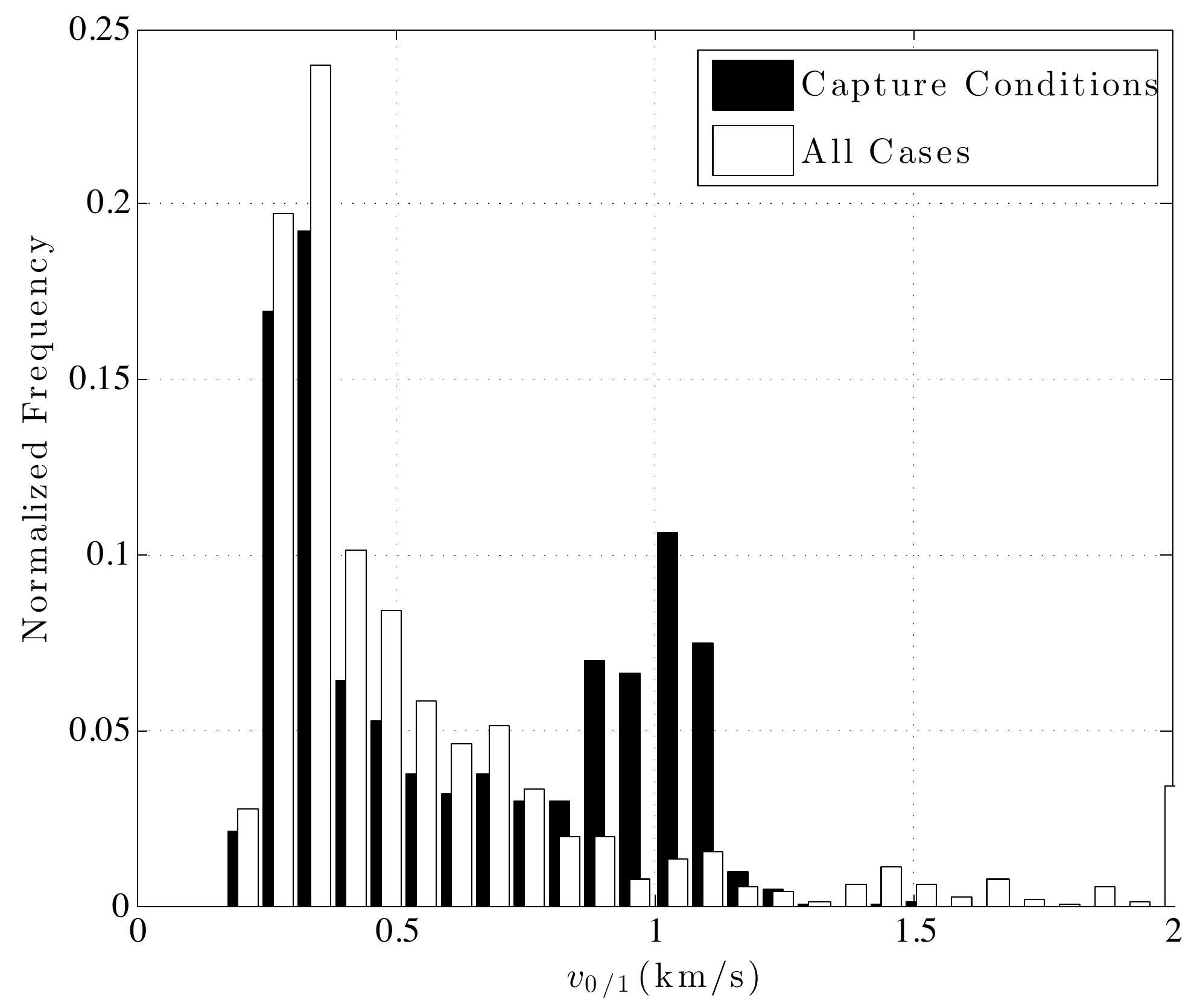}
\end{center}

{\bf Fig. 13.} Same as Figure 8 but for Case 2, with $m_1 = 1$ M$_\odot$ and $m_1^* = 0.5$ M$_\odot$.  
\label{figure:c2velP0P1}

\begin{center}
\includegraphics[scale=0.50,angle=0]{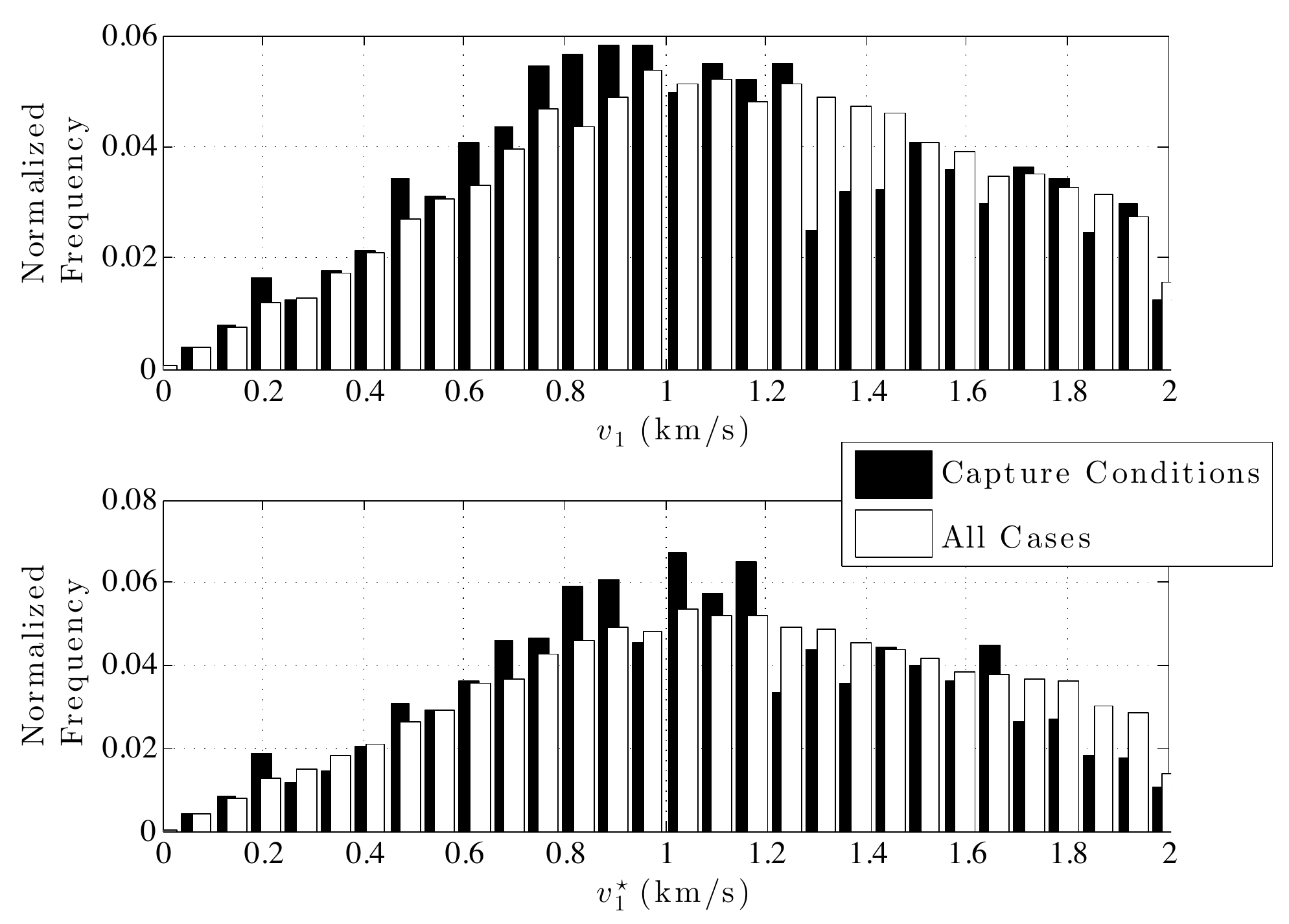}
\end{center}

{\bf Fig. 14.} Same as Figure 9 but for Case 2, with $m_1 = 1$ M$_\odot$ and $m_1^* = 0.5$ M$_\odot$. 
\label{figure:c2velP1P1}

\begin{center}
\includegraphics[scale=0.50,angle=0]{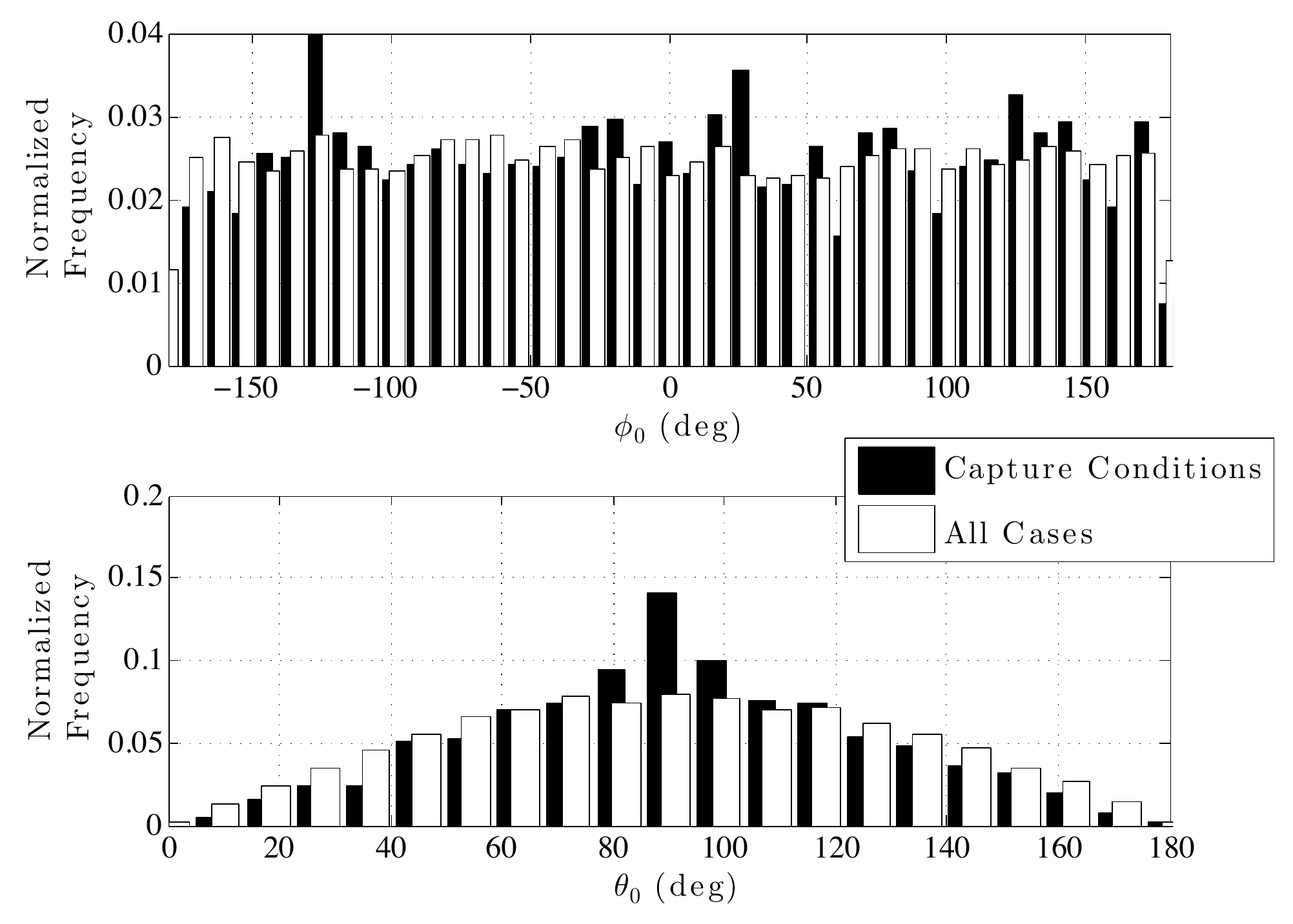}
\end{center}

{\bf Fig. 15.} Same as Figure 10 but for Case 2, with $m_1 = 1$ M$_\odot$ and $m_1^* = 0.5$ M$_\odot$. 
\label{figure:c2oriP0P1}

\begin{center}
\includegraphics[scale=0.50,angle=0]{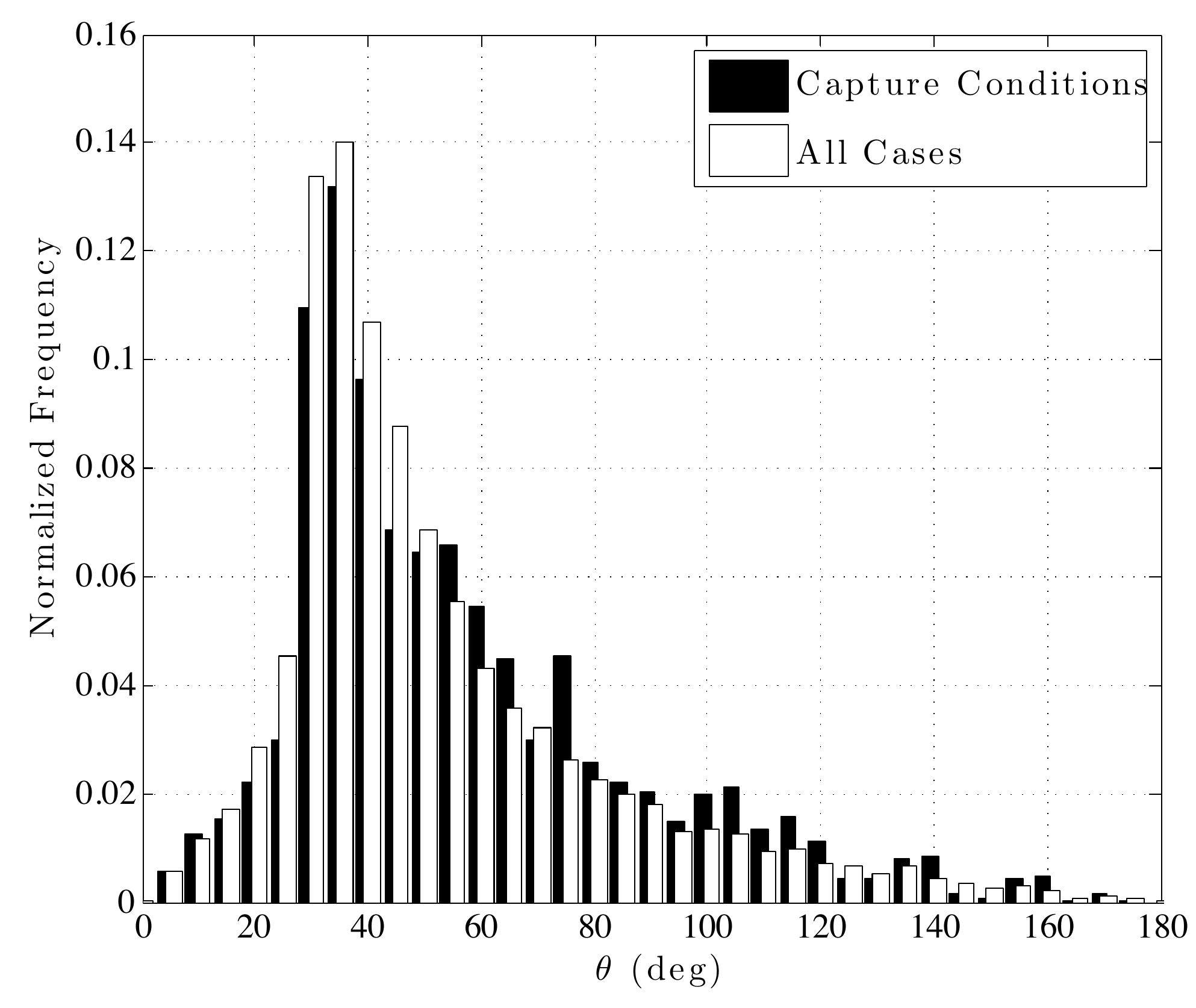}
\end{center}

{\bf Fig. 16.} Same as Figure 11 but for Case 2, with $m_1 = 1$ M$_\odot$ and $m_1^* = 0.5$ M$_\odot$.  
\label{figure:c2oriP1P1}

\begin{center}
\includegraphics[scale=0.50,angle=0]{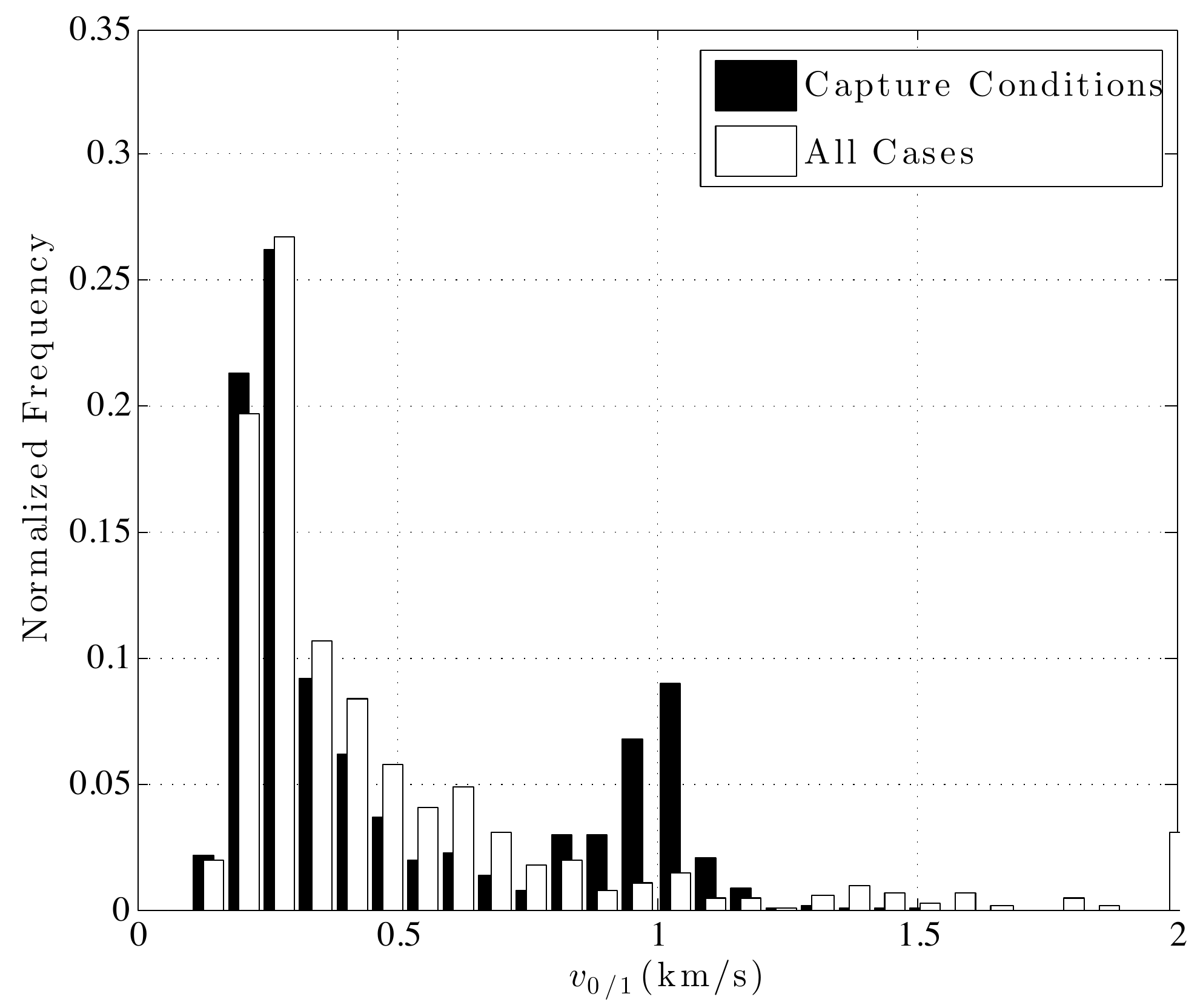}
\end{center}

{\bf Fig. 17.} Same as Figure 7 but for Case 3, with $m_1 = 0.5$ M$_\odot$ and $m_1^* = 1$ M$_\odot$. 
\label{figure:c3sepP0P1}

\begin{center}
\includegraphics[scale=0.50,angle=0]{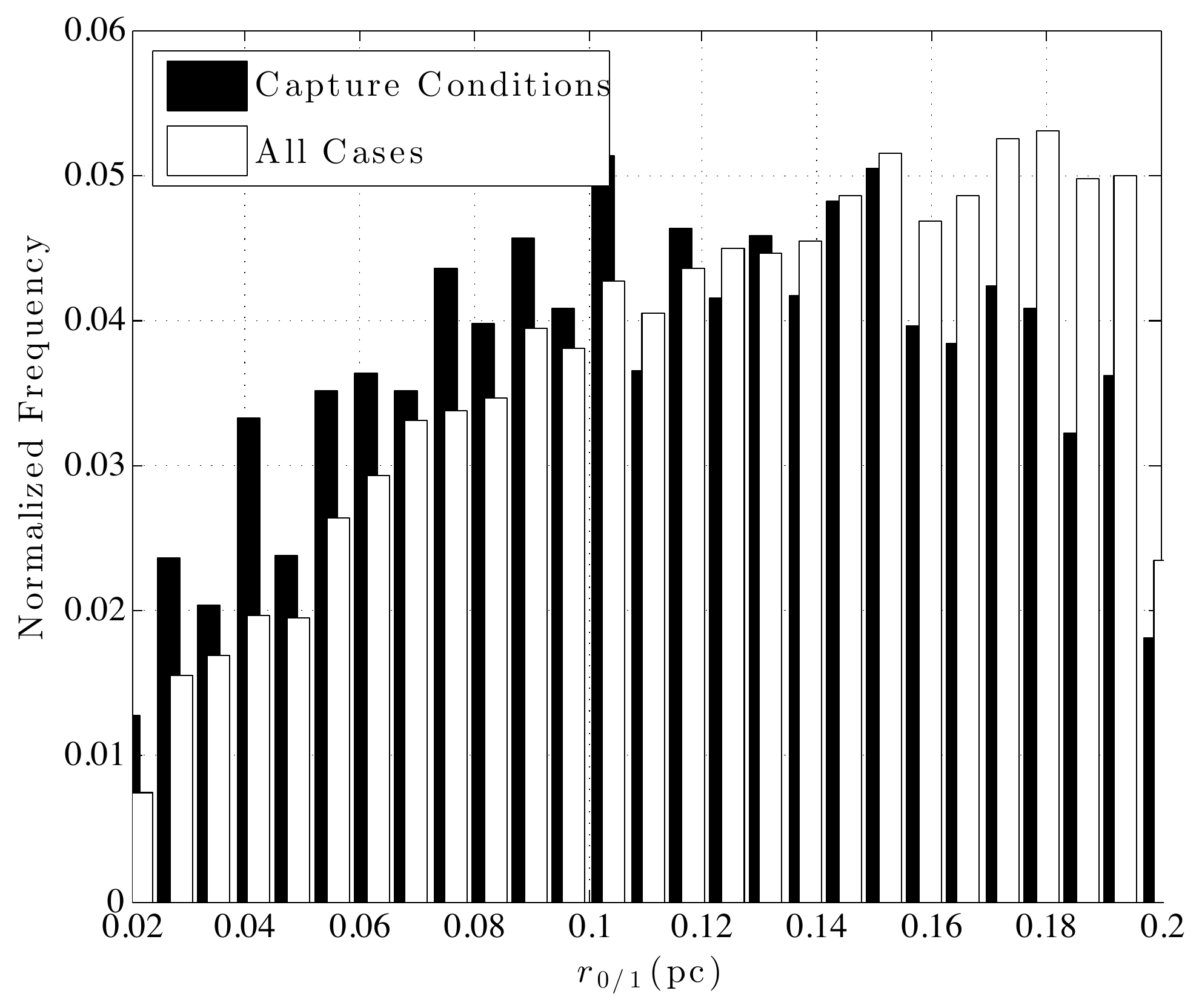}
\end{center}

{\bf Fig. 18.} Same as Figure 8 but for Case 3, with $m_1 = 0.5$ M$_\odot$ and $m_1^* = 1$ M$_\odot$. 
\label{figure:c3velP0P1}

\begin{center}
\includegraphics[scale=0.50,angle=0]{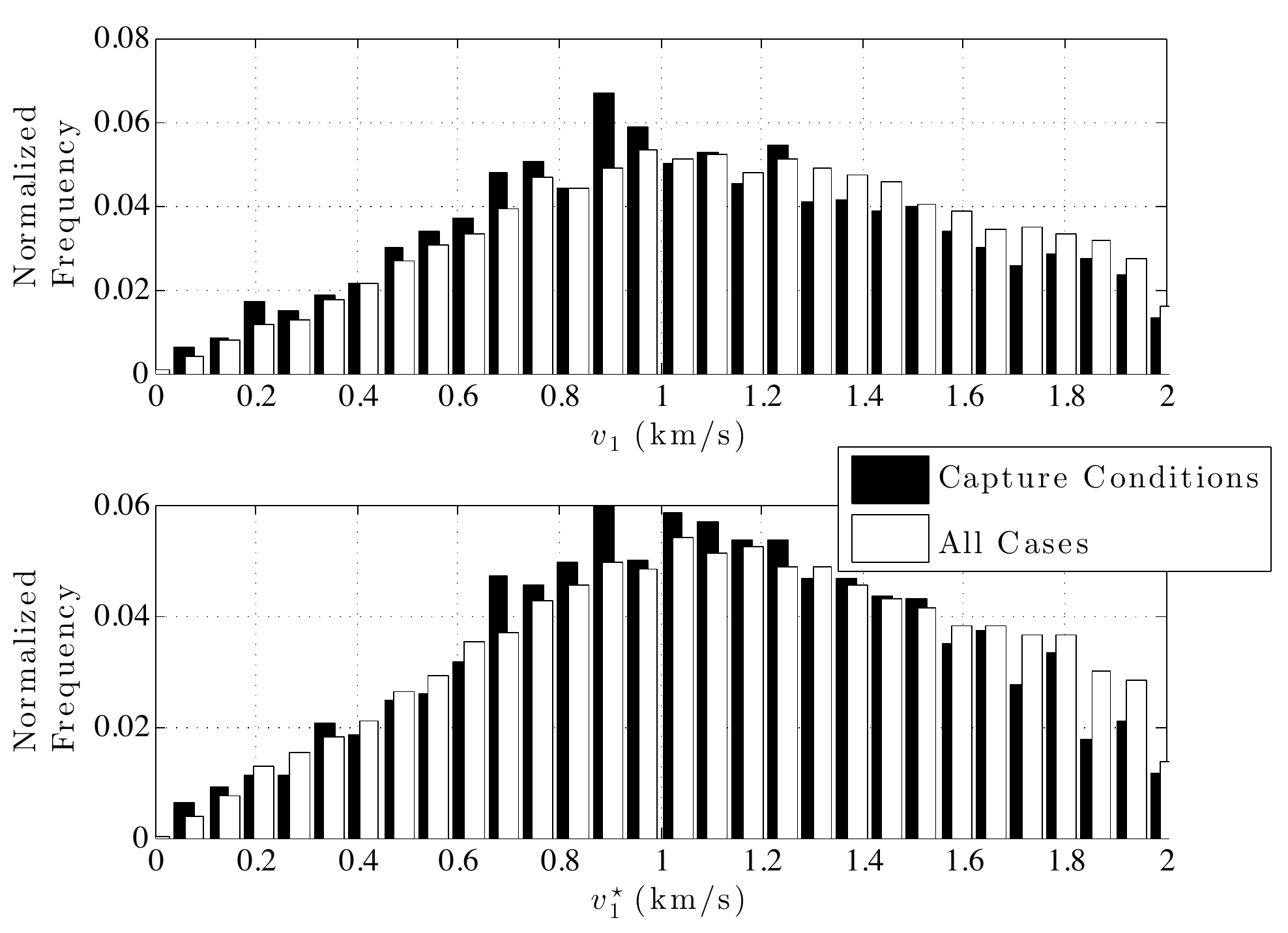}
\end{center}

{\bf Fig. 19.} Same as Figure 9 but for Case 3, with $m_1 = 0.5$ M$_\odot$ and $m_1^* = 1$ M$_\odot$.
\label{figure:c3velP1P1}

\begin{center}
\includegraphics[scale=0.50,angle=0]{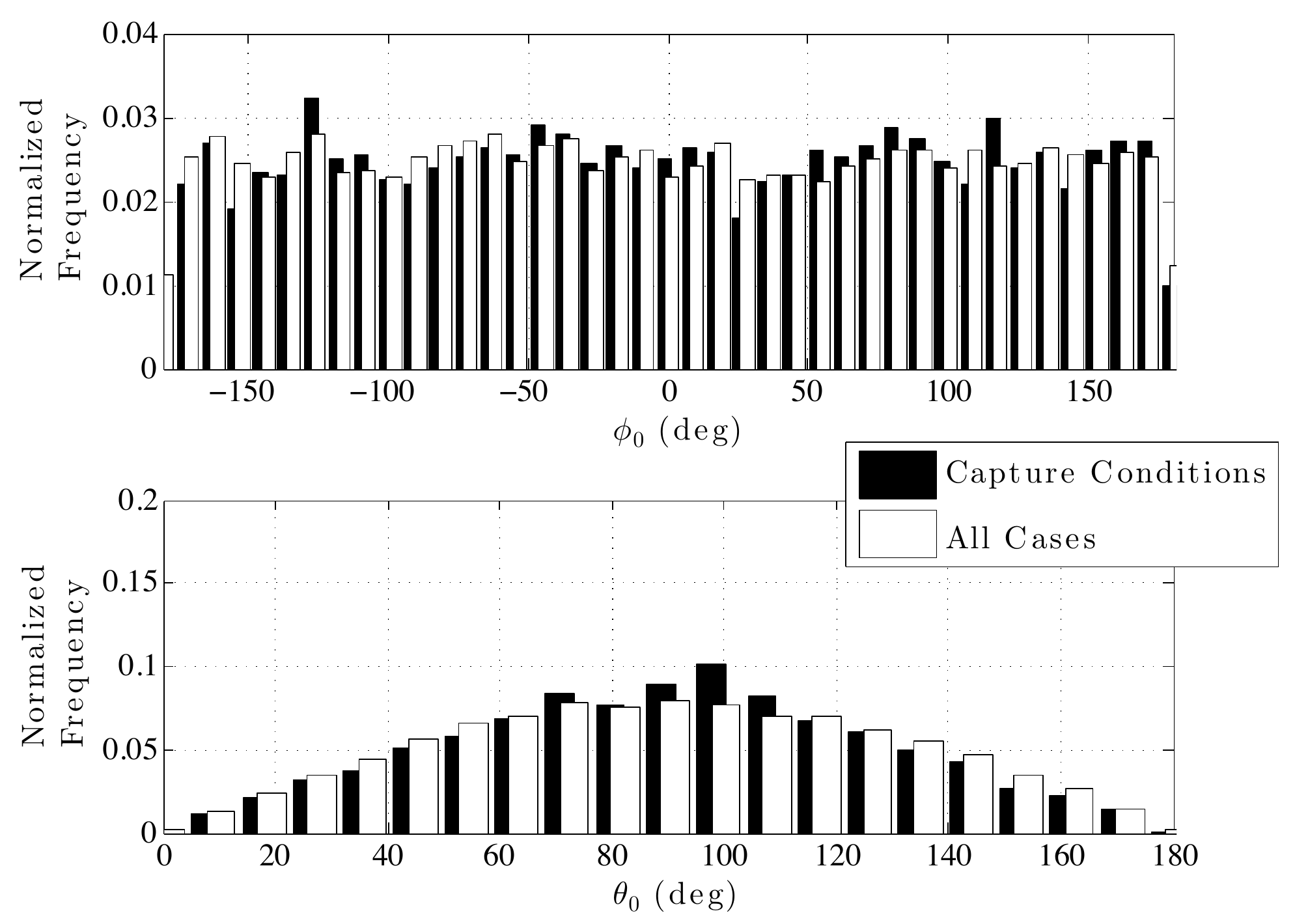}
\end{center}

{\bf Fig. 20.} Same as Figure 10 but for Case 3, with $m_1 = 0.5$ M$_\odot$ and $m_1^* = 1$ M$_\odot$.
\label{figure:c3oriP0P1}

\begin{center}
\includegraphics[scale=0.50,angle=0]{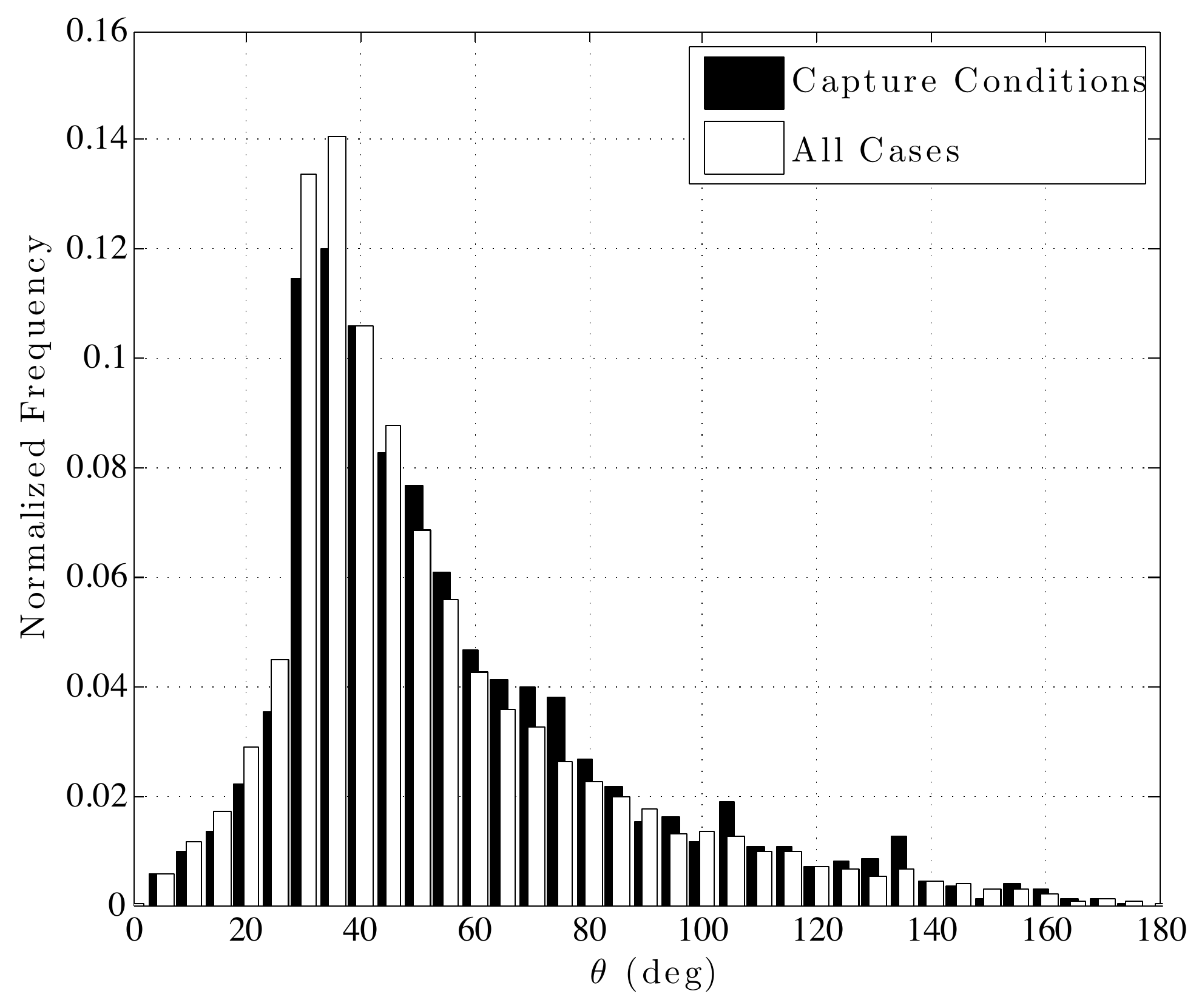}
\end{center}

{\bf Fig. 21.} Same as Figure 11 but for Case 3, with $m_1 = 0.5$ M$_\odot$ and $m_1^* = 1$ M$_\odot$.
\label{figure:c3oriP1P1}

\end{document}